\documentclass{aa}  
\usepackage{graphicx}
\usepackage{txfonts}

\begin{document} 

\newcommand\src{4U~1624--49}
\newcommand\xrism{XRISM}
\newcommand{\fetfour}{Fe~\textsc xxiv}
\newcommand{\fetfive}{Fe~\textsc xxv}
\newcommand{\fetsix}{Fe~\textsc xxvi}
\newcommand{\nitsev}{Ni~\textsc xxvii}
\newcommand{\niteig}{Ni~\textsc xxviii}

\title{A highly ionised outflow in the X-ray binary \src\ detected with XRISM}

   \author{M. D\'iaz Trigo
          \inst{1}
          \and E. Caruso\inst{2} \and E. Costantini\inst{2} \and T. Dotani\inst{3} \and T. Kohmura\inst{4} \and M. Shidatsu\inst{5} \and M. Tsujimoto\inst{3} \and T. Yoneyama\inst{3} \and J. Neilsen\inst{6} \and T. Yaqoob\inst{7}\inst{8}\inst{9} \and J. M. Miller\inst{10} 
          }

   \institute{ESO, Karl-Schwarzschild-Strasse 2, 85748, Garching bei M\"unchen, Germany
         \and
SRON Netherlands Institute for Space Research, Leiden, The Netherlands \and{Institute of Space and Astronautical Science (ISAS), Japan Aerospace Exploration Agency (JAXA), Kanagawa 252-5210, Japan} \and{Faculty of Science and Technology, Tokyo University of Science, Chiba 278-8510, Japan} \and{Department of Physics, Ehime University, Ehime 790-8577, Japan} \and{Villanova University, Department of Physics, Villanova, PA 19085, USA}\and{NASA/Goddard Space Flight Center, Greenbelt, MD 20771, USA}\and{Center for Research and Exploration in Space Science and Technology, NASA/GSFC (CRESST II), Greenbelt, MD 20771, USA} \and{Center for Space Science and Technology, University of Maryland, Baltimore County (UMBC), 1000 Hilltop Circle, Baltimore, MD 21250, USA} \and{Department of Astronomy, University of Michigan, MI 48109, USA}
             }
 
\abstract
   {The origin of accretion disc winds remains disputed to date. High inclination, dipping, neutron star Low Mass X-Ray Binaries (LMXBs) provide an excellent testbed to study the launching mechanism of such winds due to being persistently accreting and showing a nearly ubiquitous presence of highly-ionised plasmas.}
   {We aim to establish or rule out the presence of a wind in the high inclination LMXB \src, for which a highly ionised plasma has been repeatedly observed in X-ray spectra by Chandra and XMM-Newton, and a thermal-radiative pressure wind is expected.}
   {We leverage the exquisite spectral resolution of \xrism\ to perform phase-resolved spectroscopy of the full binary orbit to characterise the highly ionised plasma at all phases except during absorption dips.}
   {An outflow is clearly detected via phase-resolved spectroscopy of the source with \xrism/Resolve. Based on analysis of the radial velocity curve we determine an average velocity of $\sim$200--320~km~s$^{-1}$ and a column density above 10$^{23}$~cm$^{-2}$. The line profiles are generally narrow, spanning from $\sim$50 to ~100~km~s$^{-1}$, depending on the orbital phase, pointing to a low velocity sheer or turbulence of the highly ionised outflow and a potential increase of turbulence as the absorption dip is approached, likely due to turbulent mixing.}
   {The line profiles, together with the derived launching radius and wind velocity are consistent with a wind being launched from the outskirts of the disc and without stratification, pointing to a thermal-radiative pressure origin.}


   \maketitle
%

\section{Introduction}
\label{sec:src}

In the past two decades we have witnessed an enormous development in the studies of accretion disc winds in X-ray binaries. X-ray observations have shown the presence of highly photoionised material, revealed via absorption lines from \fetfive\ and \fetsix, in the spectra of sources at high inclination \citep{1323:boirin05aa,ionabs:diaz06aa,ponti12mnras}. Significant blueshifts have been measured in a fraction of these systems indicating outflowing material. While the outflow velocities are moderate, $\sim$300--3000 km~s$^{-1}$ \citep{mdarias25}, resulting in a modest kinetic energy output, the estimated mass outflow rates are of the order of the mass being accreted from the companion, making winds a key ingredient in the accretion process \citep{ponti12mnras,fender16lnp}. 

An important step towards understanding the effect of winds in the accretion process and their feedback to the environment is determining the wind launching mechanism. Winds can be launched via thermal, radiative or magnetic pressure but the dominant mechanism is still disputed. A fundamental prediction of thermal pressure winds is that they can only be launched at a radius larger than 0.1 of the Compton radius (R$_{IC}$), where R$_{IC}$ is the radius at which the free-fall velocity equals the isothermal sound speed of the Compton-heated gas \citep{begelman83apj}. The Compton radius is defined as R$_{IC}$\,=\,(10$^{10}$/T$_{IC,8}$)(M/M$\odot$)~cm, where T$_{IC,8}$\,=\,T$_{IC}$/10$^8$ with T$_{IC}$ being uniquely dependent on the shape of the spectrum. For X-ray binaries with a soft spectrum, T$_{IC}\sim$10$^7$~K, resulting then in R$_{IC}\sim$\,10$^{11}$\,cm for a neutron star (NS) of 1.4\,M$\odot$. 
It follows then that the velocity should be moderate, of the order of a few hundred km~s$^{-1}$, consistent with the escape velocity at such large radii. This and the fact that the flux density of the wind peaks after it is launched \citep{woods96apj,tomaru20mnras} also makes lines from thermal winds relatively narrow in the absence of additional turbulence since winds are launched across a very narrow range of radii. 

In contrast, winds launched by magnetic pressure are expected to be launched throughout the disc, resulting in wider lines and higher velocities \citep[e.g.][]{Fukumura2017}. Moreover, the lack of restriction on the launching radius implies that the capability to launch a wind does not depend on the size of the system. 

So far, winds have been found to be launched at relatively large radii and only in systems for which the Compton radius lies inside the disc \citep{diaz16an}, consistent with predictions of thermal winds. However, magnetic winds have also been invoked to explain some wind observations such as the optically thick wind of the black hole LMXB GRO~J1655-40 \citep[e.g.][]{1655:miller06nat} or the variability of such wind following changes in source luminosity or spectral hardness \citep{Neilsen2012}. Occasionally, spectra with strong absorption features from disc atmospheres show additional weaker features with relatively large blueshifts that have been interpreted as potential evidence for magnetic winds \citep{trueba20apj}.

Studying winds in NS X-ray binaries has the advantage that they have in general a known orbital period and inclination. As such, phase-resolved analyses can for example separate the contribution from the wind from absorption from other structures in the disc such as the bulge resulting from the impact of the accretion stream onto the disc. 

In this paper we leverage our knowledge of the high inclination NS X-ray binary \src\ in terms of distance, orbital period and inclination, with the new capabilities of the X-ray Imaging and Spectroscopy Mission \citep[\xrism, ][]{Tashiro25}. Launched in September 2023, \xrism\ includes a microcalorimeter onboard, Resolve, providing spectra in the 1.7-12~keV range (see \xrism\ Proposers' Observatory Guide, POG) with a resolution of $\sim$4.5~eV \citep[FWHM at 5.9~keV, ][]{porter25} and a CCD camera, Xtend, providing moderate resolution spectroscopy in the 0.4-13~keV energy range over a large region. In particular the high resolution microcalorimeter Resolve is crucial to advance our knowledge on the launching mechanism of winds and test the prediction that \src\ should be able to launch a wind via thermal pressure.

\bigskip

\src\ is a NS LMXB with a period of 20.87 hours \citep[][]{1624:smale01apj} and an inclination of 60--75$^{\circ}$, derived from the presence of absorption dips on its light curve, that are observed at the orbital period of the system, and the absence of eclipses. A distance of $\sim$15~kpc was estimated from studies of its dust scattering halo \citep{1624:xiang09apj}. The companion of \src\ was identified as a faint, Ks~=~18.3\,$\pm$\,0.1 infrared source \citep{wachter05apj} but the high extinction towards the source has prevented further characterization.

\citet{1624:parmar02aa} reported for the first time the presence of narrow absorption lines from highly ionised Fe based on XMM-Newton observations. \citet{ionabs:diaz06aa} modeled such lines with photoionised absorbers and found that spectral changes during absorption dips could be explained with an increase of the the column density of cold plasma and a decrease of ionisation of the photoionised component.   
 \citet{1624:xiang09apj} extended the work from \citet{1624:parmar02aa} by performing phase-resolved analysis of Chandra observations. They concluded that the absorption was caused by two components, one highly ionised and central, consistent with an accretion disc corona (ADC) of radius 3$\times$10$^{10}$~cm, and another one less ionised and with its ionisation parameter, $\xi$, depending on the orbital phase and consistent with being present at the disc rim, at $\sim$1.1$\times$10$^{11}$~cm. The line widths could be explained by thermal broadening, without significant turbulence nor bulk motion needed.
They also reported on a sinusoidal modulation of the flux at the orbital period, which they attributed to local cold obscuration rather than intrinsic variability of the source. 

Finally, \src\ has also been observed by the Imaging X-ray Polarimetry Explorer (IXPE). \citet{Gnarini24} reported a level of polarisation of 2.7\,$\pm$\,0.8\% during ``persistent'' emission, the highest of all neutron star LMXBs with a moderate luminosity (the so-called atoll sources), which they attributed to the Comptonised emission visible in hard X-rays and the high inclination of the system. 

\section{Observations and data reduction} 

\subsection{\xrism}

\src\ was observed twice with a four-day spacing during the Performance Verification (PV) phase of \xrism\ (see Table~\ref{tab:obslog}). 
The observations were planned to start at the beginning of the dipping phase so that each observation includes one full orbital period plus one extra dip at the end, therefore maximising the exposure time on dips while obtaining as continuous coverage as possible over one orbital period to enable phase-resolved spectroscopy. 
During the observations, Resolve was operated with the open filter.

Data were reduced with HEASoft pipeline version 6.34 and CALDB version 9 from August 2024. 
Following the recommendation of the \xrism\ ABC guide v1.0\footnote{https:\//\//heasarc.gsfc.nasa.gov\//docs\//xrism\//analysis\//abc\_guide\//xrism\_abc.html}, we performed standard additional screening based on the pulse height and rise time relation. We also set an energy threshold of 600 eV to mitigate the inclusion of crosstalk events.
Pixel 27 was not included in spectral extraction since it is known to have a significantly different gain variation compared to other pixels and can therefore degrade the spectral energy resolution. 
We applied the barycentric correction for events arrival time before extraction of data products. Only the highest resolution High primary (Hp) events were included in the final spectra. 

In-flight calibration of XRISM has shown that at high count rates, above $\sim$1~cts~s$^{-1}$~pixel$^{-1}$ for on-axis point sources, the event grade branching ratios deviate from the expected behaviour (so-called ``anomalous branching ratios'') and pixel-to-pixel spectral variations may occur.

\src\ shows a relatively constant rate along the orbital period except for the dipping phase, during which large variations are observed down to timescales of seconds. Therefore, event files have to be independently extracted for persistent and dipping phases to account for the different processing due to the high count rates of the former and the variable count rates of the latter.

In this paper, we present only the analysis of the persistent phase. During this phase, \src\ shows count rates above the limit of 1~cts~s$^{-1}$~pixel$^{-1}$ in the central four pixels of Resolve. Therefore, we next examined the event grade branching ratios for all pixels in the array. The fraction of Hp events is above 90\% for pixels in the middle and outer rings but below that value in the inner four pixels. In addition, a few of the pixels in the outer rings show a high number of Ls events, which are thought to be spurious (see XRISM POG). These spurious events introduce an uncertainty in the overall flux due to being considered as real during generation of the response and effective area files. Therefore, in what follows we excluded the pixels from the outer ring (pixels 3, 5, 6, 14, 16, 21, 23, 24, 29, 30 and 32) for spectral extraction and generation of response and effective area files to make sure that the ratio of the Hp events to the total number of events is more accurately estimated and therefore that the response and effective area assign the appropriate flux to the Hp events spectrum.

We also examined pixel-to-pixel variations of the spectra to identify any potential difference in the spectral energy dependence. Moreover, for the central pixels (pixels 0, 17, 28 and 35) we extracted time-resolved spectra per pixel and compared them with each other to assess whether pixels with count rate above the recommended limits lead to any artifacts in the data, especially related to narrow spectral features. We concluded that the per-pixel spectra were consistent with each other for each time resolved interval within the errors, and therefore that no apparent potential artifacts were present.

For the generation of response (RMF) and effective area (ARF) files we used the event screening recommended by the \xrism\ ABC guide. We specified the pixels selected for the spectra for both the RMF (pixlist) and the ARF generators (region file), excluding the outer-ring pixels, as indicated above. For extracting the RMF files we used the “L” (Large) option. In addition, for phase-resolved spectral analysis (see Sect.~\ref{sec:phase-resolved}), we used the time-filtered event files corresponding to each phase interval as input for {\tt rslmkrmf} and {\tt xaarfgen} to generate a distinct RMF and ARF for each interval.
We note that we used {\tt rslmkrmf} and {\tt xaarfgen} as opposed to {\tt rslmkrsp} since the latter was only made available after this analysis had been finalised.

We did not subtract any background since it is negligible given the high count rate of the source \citep{kilbourne2018}.

Finally, we note that the energy resolution for the Hp events has been determined to be 4.5783\,$\pm$\,0.0269~eV (FWHM) and the energy offset less than 0.088\,$\pm$\,0.022~eV for these observations based on the Mn K$\alpha$ spectrum\footnote{See https:\///heasarc.gsfc.nasa.gov\//FTP\//xrism\//postlaunch\//gainreports\//

3\//300040010\_resolve\_energy\_scale\_report.pdf and https:\///heasarc.gsfc.nasa.gov\//FTP\//xrism\//postlaunch\//gainreports\//

3\//300040020\_resolve\_energy\_scale\_report.pdf}. The achieved energy resolution and offset enables to resolve absorption lines down to widths of $\sim$50~km~s$^{-1}$ and to detect phase-dependent shifts in the energy of the lines down to a few km~s$^{-1}$ (see Sect.~\ref{sec:phase-resolved}).

\begin{table*}
\begin{center}
\caption{Observation log.}
\begin{tabular}{lllll}
\hline
\hline
   Obs.  & ID & \multicolumn{2}{l}{Date (UT)} & Exposure (ks) \\
   & & Start & End & \\
\hline
   XRISM 1  & 300040010 & 2024-03-29 23:56:18 & 2024-03-31 02:35:04 & 81 \\
   NuSTAR 1 & 30902018002 & 2024-03-30 00:01:09 & 2024-03-30 11:46:09 & 19 \\
   XRISM 2  & 300040020 & 2024-04-04 05:06:37 & 2024-04-05 07:17:04 & 64 \\
   NuSTAR 2 & 30902018004 & 2024-04-04 16:21:09 & 2024-04-05 03:16:09 & 19 \\
\hline     
\end{tabular}
\label{tab:obslog}
\end{center}
\end{table*}

\subsection{NuSTAR}

The Nuclear Spectroscopic Telescope Array (NuSTAR) observed \src\ simultaneously to XRISM in the two epochs. We extracted images, light curves and spectra for the source within a 112\arcsec\ radius circular region centred in the source and for the background within a 112\arcsec\ radius circular region 7\arcmin\ away from the source. For the extraction we used NuSTARDAS pipeline version 2.1.4 and calibration files from 20240325. 

For simultaneous fitting of Resolve and NuSTAR spectra we included a cross-normalization factor for the NuSTAR FPMA and FPMB spectra separately.

\bigskip

For spectral analysis we used {\tt spex} version 3.08.01 and {\tt xspec} version 12.15.0. Unless otherwise stated, we use optimal binning \citep{Kaastra2016} when using {\tt spex} and unbinned spectra when using {\tt xspec}. The quoted errors represent a statistical uncertainty of 1\,$\sigma$. 

\section{Analysis and results}

\subsection{Light curves}

   \begin{figure*}
   \includegraphics[width=1.\textwidth]{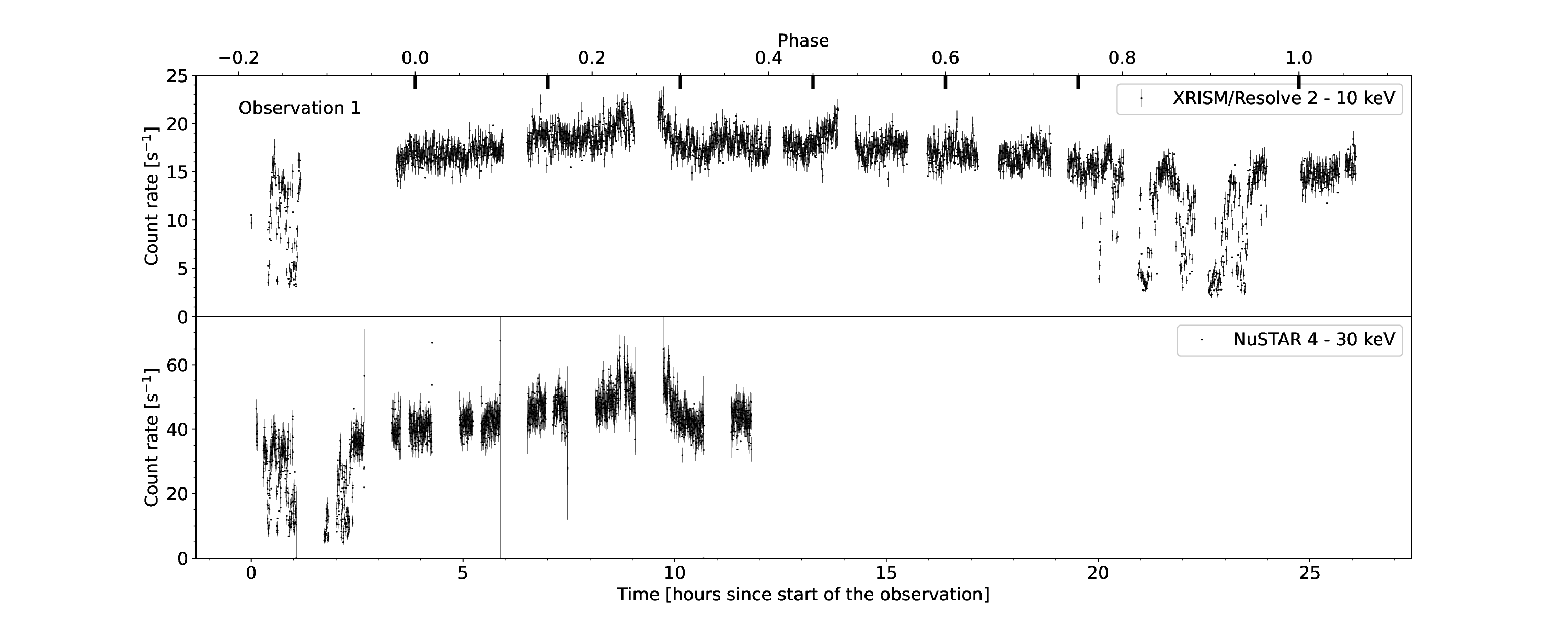}
   \includegraphics[width=1.\textwidth]{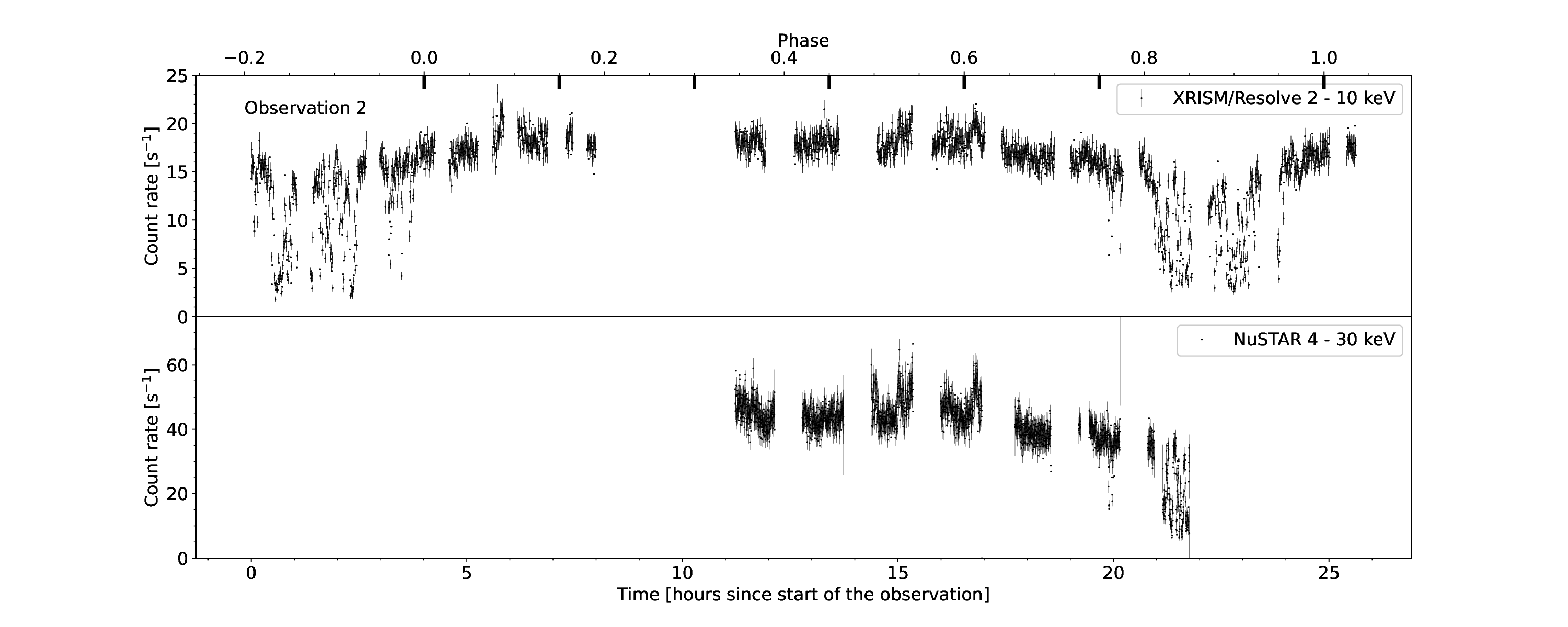}
   \caption{XRISM/Resolve and NuSTAR light curves in the 2--10~keV and 4--30~keV energy range, respectively, for observations 1 (top) and 2 (bottom) showing absorption dips at the start and end of the observations. The binning is 25 s in each panel. The thick tick marks at the top axis mark the start and end of phase-resolved intervals.}
    \label{lc}%
    \end{figure*}
    
Fig.~\ref{lc} shows the light curves of the two \xrism\ observations in the 2--10~keV energy range and of the two NuSTAR observations in the 4--30~keV energy range. 
The light curves are similar in the two epochs. Each \xrism\ observation shows two full dip events, at the start and the end of the observation. Outside of the dips,  hereafter so-called ``persistent'' emission, the curves show significant orbital modulation, with the flux raising after a dipping episode until it reaches a maximum at the opposite phase of the dips and declining thereafter until the next dipping episode is reached. 

The NuSTAR observations are simultaneous to the \xrism\ observations, but each observation covers only half the orbital period of the system. However, the availability of NuSTAR spectra for all orbital phases at least from one observation and the similarity of the two observations, as indicated by the XRISM light curves, makes it possible to perform broadband spectral fits.

In this paper, we concentrate in the analysis of the ``persistent'' emission, and we refer to Caruso et al. (in prep) for a detailed analysis of the dips. 

\subsection{Spectral analysis}
\label{sec:spectral}

To inform our detailed analysis and determine the general properties of the plasma, we first extracted an aggregated spectrum for the persistent emission for each observation.

We found that the continuum from the \xrism/Resolve spectra could be satisfactorily fitted with a blackbody component ({\tt bb} component in {\tt spex}) modified by cold absorption from the interstellar medium ({\tt hot} component in {\tt spex}). However, the hard X-ray spectra available from the NuSTAR observations indicate that a second component is necessary to fit the hard X-rays. Therefore, we added to the continuum a second component consisting of Comptonisation of soft photons in a hot plasma using the blackbody as the source for the seed photons ({\tt comt} component in {\tt spex}). Thus, our Model~1 is {\tt hot*(bb+comt)} in {\tt spex} notation.

 Similar to previous observations from XMM-Newton and Chandra \citep{1624:parmar02aa, 1624:xiang09apj}, the residuals show clear narrow and deep absorption features from highly ionised Fe (see Fig.~\ref{narrow_lines}). The unprecedented spectral resolution of \xrism/Resolve reveals for the first time a clear blueshift in the \fetfive\ and \fetsix\ lines with respect to their rest energy. In addition, the spin-orbit doublet of \fetsix\ is resolved into two distinct, narrow, features. 

 Given the excellent statistics of the spectra and the orbital modulation apparent in the light curves we chose not to perform a detailed analysis of the aggregated spectra and continue instead with phase-resolved analysis in the remainder of this paper.

 We note that previous spectral analyses of \src\ with XMM-Newton and Chandra \citep{ionabs:diaz06aa,1624:xiang07apj} included the contribution of a dust scattering halo. This halo is produced by a dust cloud in the line of sight to the source, at a distance of approximately 15~kpc \citep{1624:xiang07apj}, which scatters light from the source and delays the arrival of its emission by 1.6~ks seconds. Based on Fig.~7 from \citet{1624:xiang07apj}, the contribution of the scattering halo is significant up to $\sim$1$\arcmin$ in radius and decreases thereafter. Given that the PSF of XRISM, 1.3$\arcmin$, is significantly larger than that of Chandra, 0.5$\arcsec$, or XMM-Newton, 5$\arcsec$, we estimate that inclusion of a dust scattering halo component is not needed for analysis of the persistent phase spectra. 
 
 However, we note that this may not be the case when analysing the dipping spectra. In such a case, the time delay introduced by the scattering will result in emission from the persistent intervals scattered in during dipping intervals and such scattered emission from the persistent intervals may be significant compared to the low flux characteristic of dipping intervals. The larger Field of View of Xtend when compared with Resolve may also be helpful to fully characterise the halo contribution during that interval. We refer to Caruso et al. (in prep) for further discussion on this topic.

\subsection{Phased-resolved spectral analysis}
\label{sec:phase-resolved}

We next used the orbital period of the source of 20.87~h to slice the persistent emission in five intervals, each covering 0.15 in orbital phase (p0--p4, with p0 starting at phase 0 and p4 ending at phase 0.75), thus for a total of three quarters of the full orbit. We used the same ephemeris (2024-03-30 03:57:15~UT as the start time of p0 for Obs~1) for the two observations to make sure that the same orbital phases are covered. We define $\varphi_0$ such that intervals p0-p4 do not include episodes of deep dipping in any of the observations (see Fig.~\ref{lc}). This is to avoid the low ionisation plasma present during dips and associated to the accretion stream or disc bulge \citep{1323:boirin05aa, ionabs:diaz06aa} affecting our characterisation of the highly ionised plasma.  

We also note that inspection of the continuum parameters and residuals for the phase-resolved intervals indicates that observations 1 and 2 have similar spectra within the uncertainties. Therefore, we merged the event files of the two observations to increase the statistics and re-extracted the phase-resolved spectra for the merged event file, which we use in the remainder of this paper.

\subsubsection{Continuum fits}

\begin{figure}
\centering

\includegraphics[width=0.48\textwidth]{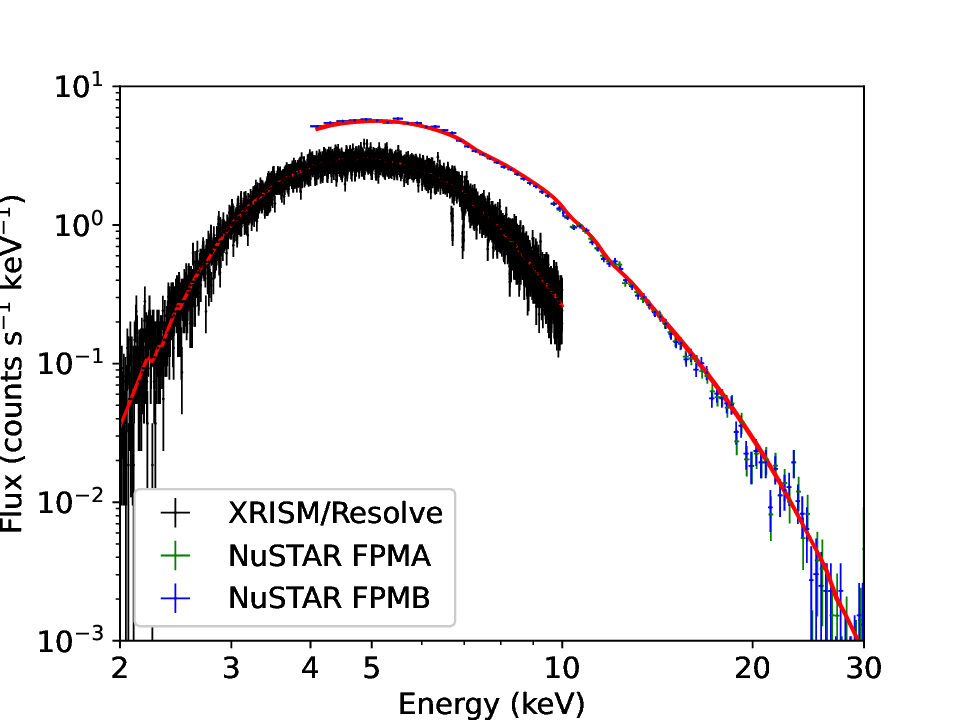}

\caption{\xrism/Resolve and NuSTAR spectra for interval p2 fitted with Model 1 ({\tt hot*(bb+compt)}, see text)}. 

\label{broadband_fit}
\end{figure}

\begin{figure}
\centering
\includegraphics[width=0.5\textwidth]{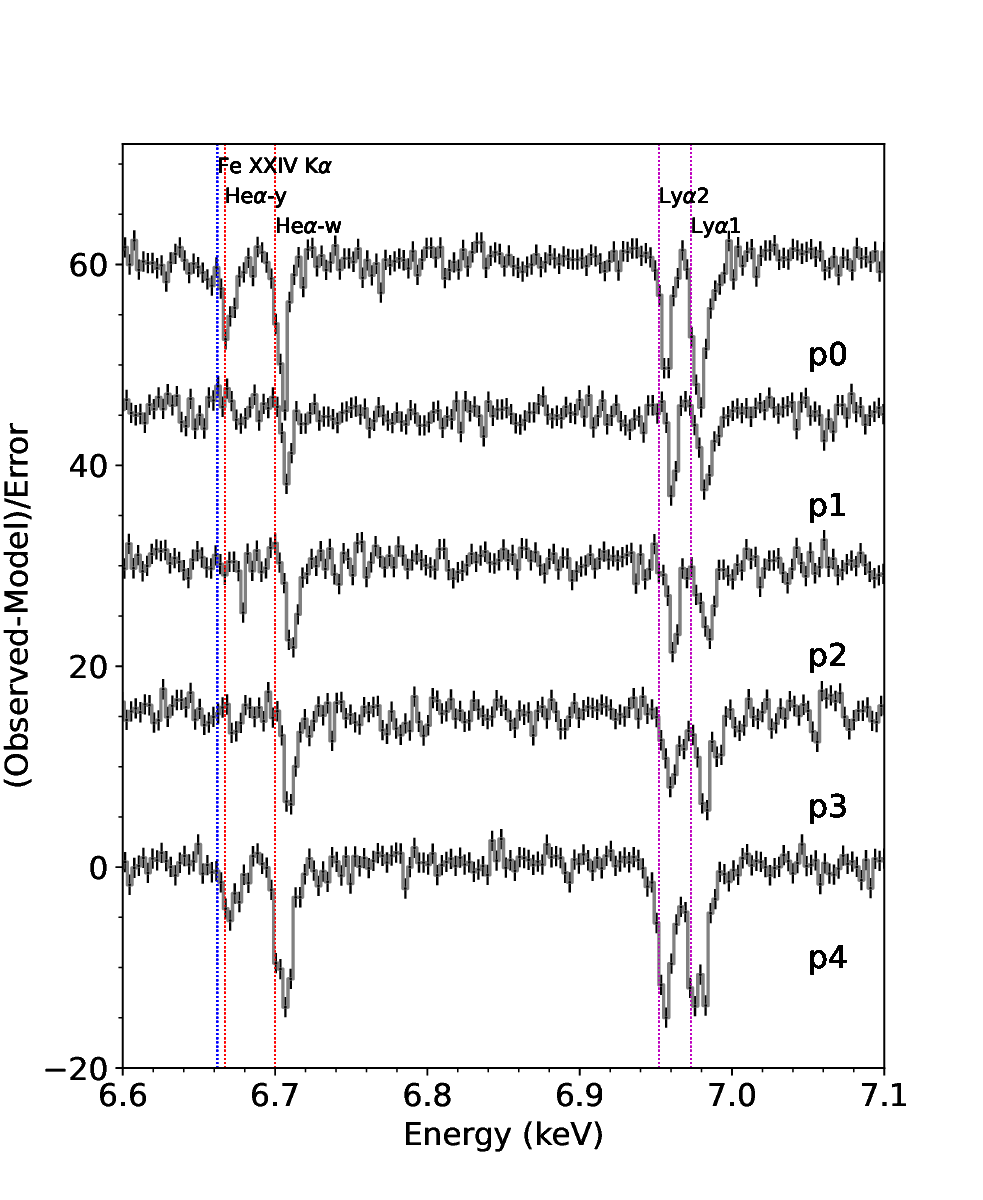}
\caption{Residuals in the Fe~K$\alpha$ energy range for intervals p0--p4 after fitting of the continuum. The y-scale has been shifted for p0-p3 for clarity.    
}
\label{narrow_lines}%
\end{figure}

We first re-fitted the \xrism\ and NuSTAR spectra for each of the phase-resolved intervals (after merging the two observations) with the same continuum as that used for the phase-aggregated spectrum (Model 1). 
Despite having broadband spectra, we found that the opacity of the Comptonisation component is unconstrained in the fits. Therefore, we fixed the opacity to a value of 10, which results in values for the temperature of the seed photons and of the electrons similar to those recently obtained by \citet{Gnarini24}. 

This model fits the spectra of the five intervals well. As an example, Fig.~\ref{broadband_fit} shows the broadband fit for interval p2, with clear narrow residuals at the Fe~K energy region. 
A more detailed look to the residuals in the Fe~K$\alpha$ region for all intervals (Fig.~\ref{narrow_lines}) shows additional important differences among intervals. First, blueshifts are apparent for all lines with respect to their rest energy, but the blueshifts differ among the intervals, with p0 and p4 showing the smallest blueshift and p2 the largest. 

A prominent feature at $\sim$6.66~keV is present only in intervals p0 and p4, most likely due to the presence of the intercombination line of \fetfive\ or \fetfour. In addition, the Ly$\alpha$1 line shows an apparent blue wing in p0 and a double peak, reminiscent of two components with different velocities, in p4.  

The larger column density of \fetsix\ in intervals p0 and p4 is confirmed by the presence of transitions up to Ly$\gamma$ in those intervals. This larger column density could be caused by a plasma with lower ionisation 
for which ions that were fully stripped in intervals p1--p3 have now gained an electron and become visible. 

Finally, the ratio of the lines in the \fetsix\ spin-orbit doublet shows some differences among intervals. This can point out to different solid angles and fraction of scattered emission of the plasma. However, the fact that Ly$\alpha$2 is deeper than Ly$\alpha$1 for interval p2 (against the expected 1:2 ratio for Ly$\alpha$2: Ly$\alpha$1) may be also pointing to saturation effects derived from large column densities.

\subsubsection{Fits to individual species}

To evaluate the effect of saturation on the lines in the fits with photoionised plasmas, we first performed fits to \fetfive\ and \fetsix\ lines separately and for two different energy ranges: the 6.5--9~keV energy range, which includes all important transitions, and the 7.5--9~keV energy range, which does not include the K$\alpha$ (He$\alpha$ for \fetfive\ or Ly$\alpha$ for \fetsix) components that we suspect to be saturated.

For this, we used the {\tt slab} component in {\tt spex}, in which the column density of each species can be fitted independently taking the width and shift of lines into account. This component treats in a simplified manner the absorption by a thin slab composed of different ions and is a good approximation as long as the slab is not too large \citep{Kaastra2002}. 
To improve the measurement of the line shifts, we also used the \nitsev\ and \niteig\ species that we do not expect to be saturated. Due to the very narrow energy range used for this test, we used a simplified continuum of a blackbody with temperature and normalisation as free parameters for each interval. Table~\ref{tab:slabspecfits} 
shows the results of the fits for all intervals for the 7.5--9~keV energy range. 

\begin{table*}
    \centering
    \caption{Spectral fits of the \fetfive, \fetsix, \nitsev\ and \niteig\ species for p0-p4 intervals with the {\tt slab} component (see text). For each interval we fit only the \xrism/Resolve spectra in the 7.5--9~keV energy range.}
    \begin{tabular}{lccccc}
        \hline
         Interval & p0 & p1 & p2 & p3 & p4 \\
         \hline
         \bigskip
         Parameter & \\
         \smallskip
         log\,($N_{\rm \fetfive}\,\,{\rm cm^{-2}}$) & 17.63\,$\pm$\,0.07 & 17.15\,$^{+0.20}_{-0.27}$ & 17.34\,$^{+0.11}_{-0.15}$ & 17.43\,$^{+0.10}_{-0.12}$ &  17.74\,$\pm$\,0.07\\

         log\,($N_{\rm \fetsix}\,\,{\rm cm^{-2}}$) & 18.15$\pm$0.06 & 17.93$\pm$0.14 &	17.98$\pm$0.09 & 17.99$\pm$0.09 & 18.18$\pm$0.06 \\
         log\,($N_{\rm \nitsev}\,\,{\rm cm^{-2}}$)  & 16.70$^{+0.11}_{-0.14}$ & -- & -- & 16.32$^{+0.23}_{-0.46}$ & 16.93$^{+0.08}_{-0.10}$\\
         log\,($N_{\rm \niteig}\,\,{\rm cm^{-2}}$)  & 16.86\,$^{+0.15}_{-0.23}$ & 16.6\,$^{+0.3}_{-0.6}$ & 17.02$^{+0.14}_{-0.19}$ & 16.87$^{+0.17}_{-0.26}$ & 17.02$^{+0.13}_{-0.17}$\\
         \smallskip\\
        $\sigma_v$ ({$\rm km~s^{-1}$}) & 130\,$\pm$\,25 & 45\,$^{+30}_{-44}$ & 116\,$\pm$\,36 & 104\,$^{+29}_{-26}$ & 120\,$\pm$\,26 \\
         v ({$\rm km~s^{-1}$}) & -150\,$\pm$\,19 & -405\,$\pm$\,28 & -456\,$\pm$\,30 & -377\,$\pm$\,27 & -194\,$\pm$\,17 \\
         
         \hline\\
    \end{tabular}
    \label{tab:slabspecfits}
\end{table*}

When the K$\alpha$ transitions are included in the fit, the fitted column densities are lower and the K$\beta$ transitions are not well fitted, indicating line saturation that is not being accounted for with this simple model, likely because of the presence of scattered radiation not allowing the absorption lines reach zero flux. 
Because of this saturation effect, we first focus in the values obtained from the fits to the 7.5--9 keV energy range, which allow to get a first estimate of the column densities and velocities with this simple model. 

An increase of column density of both \fetfive\ and \fetsix\ is observed in intervals p0 and p4, as expected from Fig.~\ref{narrow_lines}. In addition, a clear modulation is observed in the shift of the lines, with a maximum of $\sim$450~km s$^{-1}$ observed during p2. 

We note that we also attempted to fit each species separately. However, due to the larger uncertainties of the width and shift of the lines in the fits to the individual species, we cannot make any definite conclusion at this stage on whether the location of the individual species is different. Therefore, we only show the joint fit to the Fe~K$\beta$, \nitsev\ and \niteig\ lines, which allows to reduce the uncertainties in the velocity shift compared to those obtained fitting each species individually. 

\subsubsection{Fits with photoionised models}
\label{pionfits}

\begin{figure*}

\includegraphics[width=1.\textwidth,angle=0]{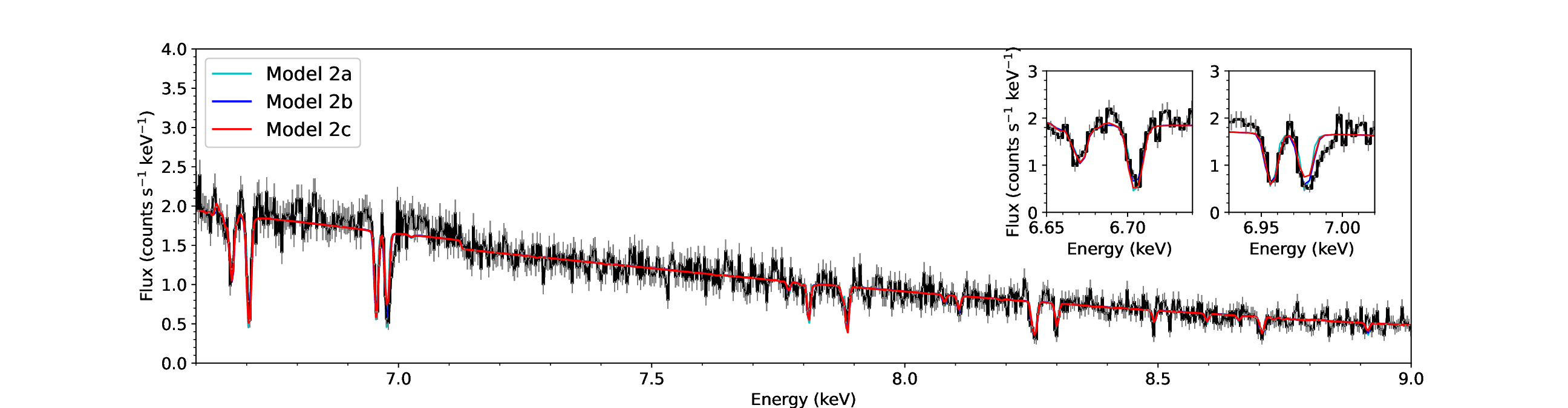}
\includegraphics[width=1.\textwidth,angle=0]{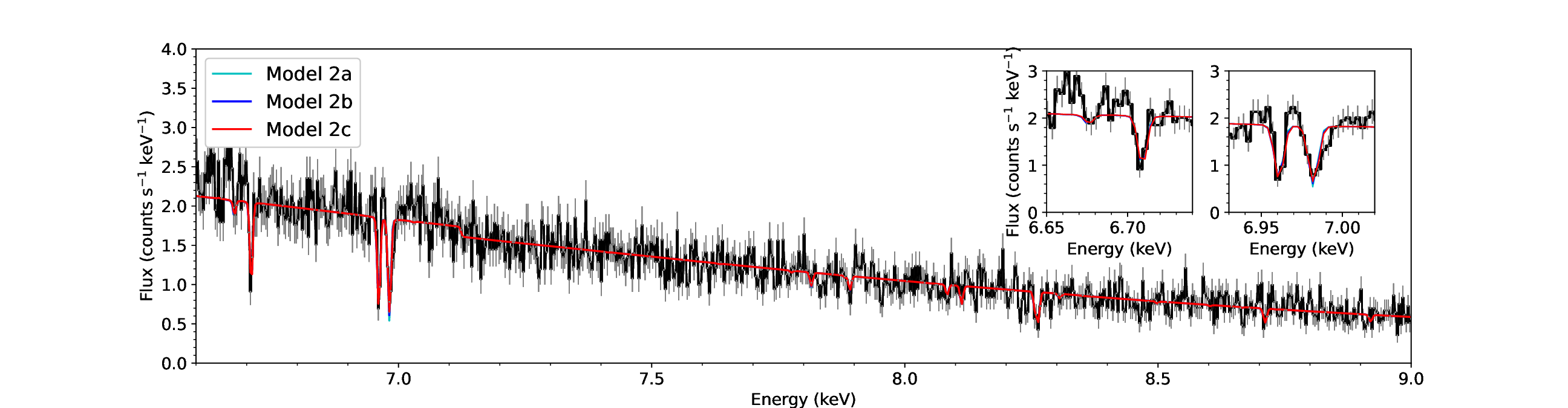}
\includegraphics[width=1.\textwidth,angle=0]{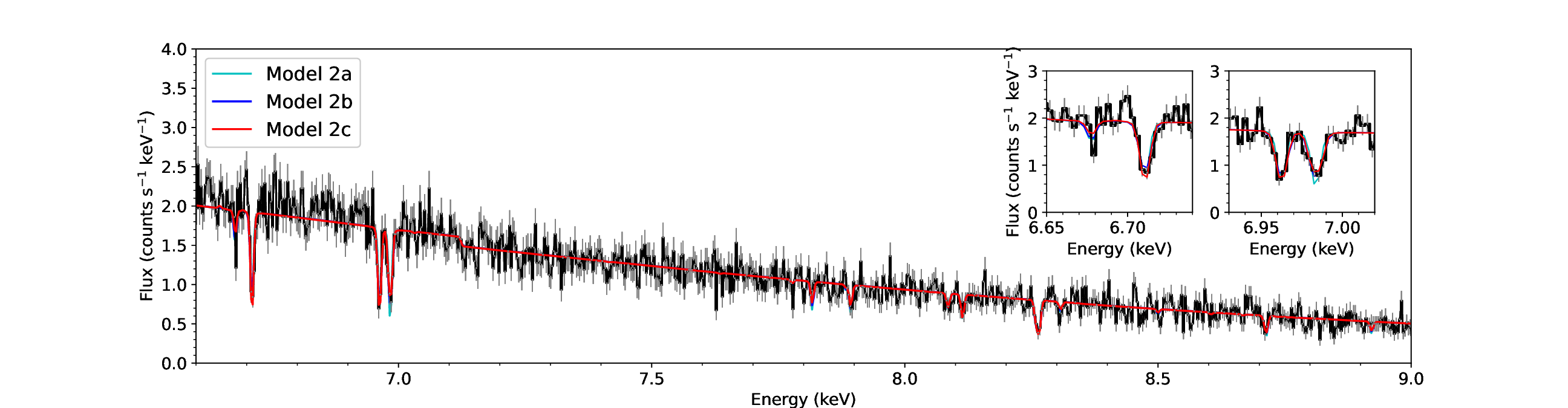}
\includegraphics[width=1.\textwidth,angle=0]{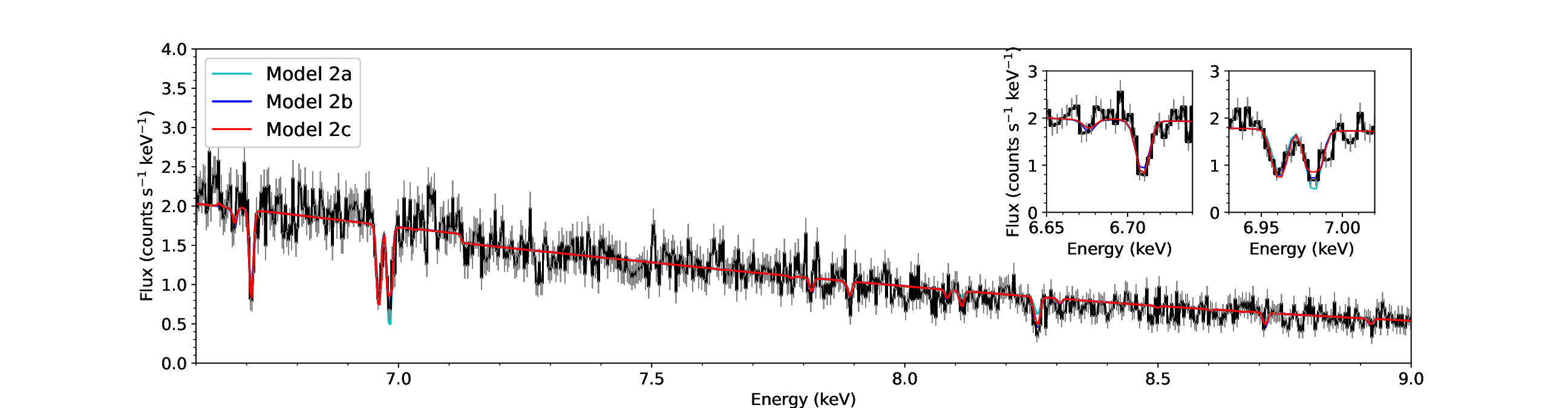}
\includegraphics[width=1.\textwidth,angle=0]{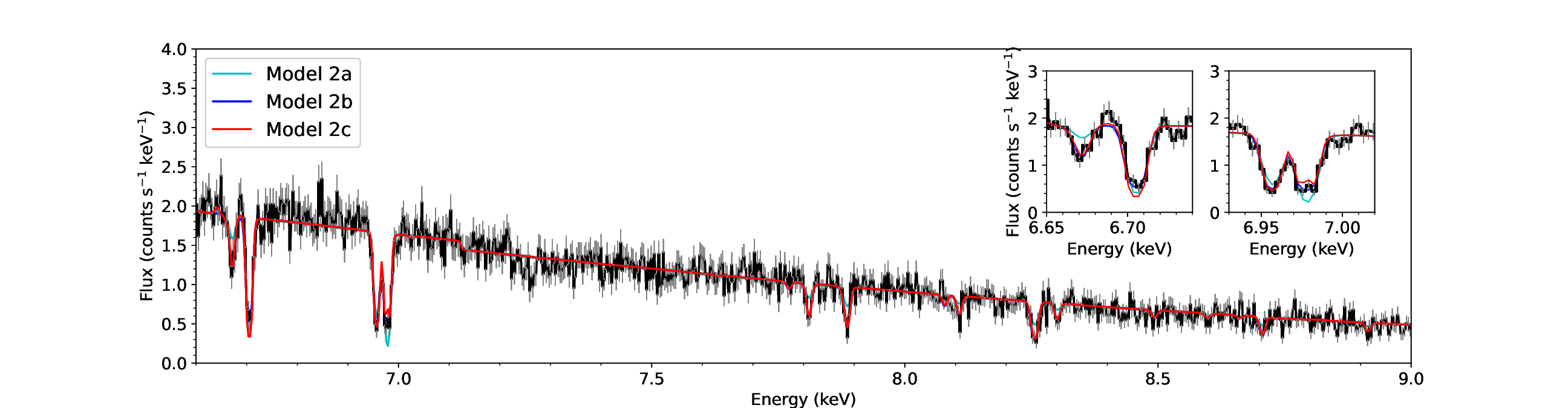}
\caption{Spectra for intervals p0 to p4 (top to bottom, respectively) with models 2a, 2b and 2c overplotted in cyan, blue and red, respectively (see Tables~\ref{tab:specfits1b}--\ref{tab:specfits1c} for fitted parameters). The insets show a detail of the \fetfive\ and \fetsix\ K$\alpha$ regions highlighting the different prediction of the three models for the \fetsix\ spin-orbit doublet ratio.} 
\label{fig:fcov_vs_omega}
\end{figure*}

To provide a more detailed characterisation of the ionised plasma, we next substituted the {\tt slab} component by the {\tt pion} component in {\tt spex} \citep{Mehdipour16}. This component uses the ionising radiation from the continuum component self-consistently, so that the ionisation balance and the spectrum of the photoionised plasma always corresponds to the fitted continuum and is particularly adequate when broadband continua are available, as it is the case in this paper. 
We first allowed the column density, ionisation parameter, velocity width and velocity shift of the {\tt pion} component to vary ({\tt hot*pion$_1$*(bb+comt) in spex}, hereafter Model~2a). This model significantly improves the fit compared to Model 1 for all intervals ($\Delta\,C-stat$ between 183 and 930 for 4 additional d.o.f), with the largest improvement observed for intervals p0 and p4. However, as indicated before, the line ratio of the \fetsix\ spin-orbit doublet differs from the expected 2:1 ratio in all intervals and even inverts in p1, p2 and p4. Therefore, we also tried variations of Model 2a leaving the covering factor (Model 2b) or the solid angle (Model 2c) as free parameters in the {\tt pion} component. 
A covering factor less than one implies that the ``uncovered''  continuum emission does not undergo absorption by the photoionised plasma, and therefore mimics scattered continuum by the photoionised plasma. In contrast, a solid angle larger than zero includes emission from the absorbing plasma but does not include the scattered continuum. In particular, in Model~2c the emitting plasma is forced to have the same parameters (column density, ionisation parameter, velocity broadening and velocity shift) than the absorbing plasma. This implies that we only include emission from the plasma that is also absorbing, and not from plasma that is e.g. at other orbital phases and we cannot see via absorption. 
We find that the two modified models improve the fits in general. For p4, the improvement is large,  with $\Delta\,C-stat$ of 138/108 for Models 2b/2c for one additional d.o.f, respectively. For intervals p0-p3 the improvement is less prominent with $\Delta\,C-stat$ between $\sim$12 and 45 for one additional d.o.f. The results of the fit for Models 2b and 2c are shown in Tables~\ref{tab:specfits1b} and \ref{tab:specfits1c}, respectively. Fig.~\ref{fig:fcov_vs_omega} shows the fit with the different models in the relevant Fe~K region for all intervals.

\begin{table*}
    \centering
        \caption{Spectral fits of the p0-p4 intervals with Model 2b (see text). For each interval we fit the 2--10~keV XRISM and 4--30~keV NuSTAR spectra simultaneously. kT$_{bb}$ is coupled for the {\tt blackbody} and {\tt comt} components. $f$ indicates a fixed parameter and $c$ a coupled parameter. The cross-normalization for NuSTAR with respect to Resolve is indicated in the last rows. We note that NuSTAR spectra were available only for one observation for all intervals except p2, for which we fitted the spectra from the two observations separately and therefore indicate two normalizations.}

    \begin{tabular}{lccccc}
        \hline
         Interval & p0 & p1 & p2 & p3 & p4 \\
         \hline
         \smallskip
         Parameter & \\
         & \multicolumn{5}{c}{\em hot} \\
         \smallskip
         N$_H^{cold}$ (10$^{22}$ cm$^{-2}$) & 6.0$\pm$0.1 & 6.3$\pm$0.1 & 6.4$\pm$0.1 & 5.9$\pm$0.1 & 5.9$\pm$0.1\\
         \smallskip
         & \multicolumn{5}{c}{\em blackbody} \\
         \smallskip
         kT$_{bb}$ (keV) & 1.23$\pm$0.01 & 1.11$\pm$0.02 & 1.18$\pm$0.01 & 1.18$\pm$0.01 & 1.19$\pm$0.01 \\
         Norm (10$^{13}$ cm$^2$) & 0.97$\pm$0.04 & 0.2$\pm$0.1 & 1.00$\pm$0.06 & 0.73$\pm$0.06 & 0.88$\pm$0.05 \\
         \smallskip
         & \multicolumn{5}{c}{\em comt} \\
         \smallskip
         kT$_{bb}$ (keV) & (c) & (c) & (c) & (c) & (c) \\
         kT$_{hot}$ (keV)& 2.92$\pm$0.04 & 2.92$\pm$0.02 & 2.90$\pm$0.03 & 2.97$\pm$0.03 & 2.91$\pm$0.04 \\
         $\tau$ & 10 (f) & 10 (f) & 10 (f) & 10 (f) & 10 (f)  \\
         Norm (10$^{44}$ ph~s$^{-1}$~keV$^{-1}$)& 23$\pm$2 & 41$\pm$3 & 27$\pm$3 & 30$\pm$2 & 27$\pm$2 \\
         \smallskip
         & \multicolumn{5}{c}{\em pion$_1$} \\
         \smallskip
         N$_H^{pion}$ (10$^{22}$ cm$^{-2}$) & 18$\pm$3 & 12$^{+6}_{-4}$ & 30$^{+13}_{-8}$ & 24$^{+10}_{-6}$ & 25$\pm$4 \\
         log $\xi$ (erg~cm~s$^{-1}$) & 3.64$\pm$0.04 & 3.97$\pm$0.09 & 3.97$\pm$0.07 & 3.95$\pm$0.07 & 3.76$\pm$0.04 \\
         $\sigma_v$ (km~s$^{-1}$) & 80$\pm$7 & 49$^{+15}_{-18}$ & 71$\pm$11 & 111$^{+18}_{-15}$ & 150$\pm$7 \\
         v (km~s$^{-1}$) & -196$\pm$8 & -395$\pm$13 & -461$\pm$12 & -392$^{+14}_{-17}$ & -218$\pm$10 \\         
         Covering fraction & 0.71$\pm$0.03 & 0.83$^{+0.14}_{-0.09}$ & 0.65$\pm$0.04 & 0.65$\pm$0.04 & 0.76$\pm$0.02 \\
        \hline\\
         Lumin$_{0.013-13.6~keV}^{total}$ (erg~s$^{-1}$) & 5.7$\times$10$^{37}$ & 6.1$\times$10$^{37}$ & 6.0$\times$10$^{37}$ & 5.9$\times$10$^{37}$ & 5.8$\times$10$^{37}$\smallskip\\
          \hline \\
         $Cstat$ (d.o.f.) & 3238 (2924) & 2946 (2726) & 3431 (3021) & 2994 (2840) & 2976 (2861) \\
        $\Delta Cstat$ (d.o.f.) (compared to Model 2a) & -43 (1) & 0 (1) & -15 (1) & -28 (1) & -135 (1) \\
         \hline\\
         Cross-normalization NuSTAR FPMA & 0.99 & 1.01 & 1.00/1.08 & 1.01 & 1.00 \\
         Cross-normalization NuSTAR FPMB & 0.98 & 1.00 & 1.03/1.02 & 1.00 & 0.99 \\
         \hline\\
    \end{tabular}
    \label{tab:specfits1b}
\end{table*}

\begin{table*}
    \centering
        \caption{Spectral fits of the p0-p4 intervals with Model 2c (see text). For each interval we fit the 2--10~keV XRISM and 4--30~keV NuSTAR spectra simultaneously. kT$_{bb}$ is coupled for the {\tt blackbody} and {\tt comt} components. $f$ indicates a fixed parameter and $c$ a coupled parameter. The cross-normalization for NuSTAR with respect to Resolve is indicated in the last rows. We note that NuSTAR spectra were available only for one observation for all intervals except p2, for which we fitted the spectra from the two observations separately and therefore indicate two normalizations.}

    \begin{tabular}{lccccc}
        \hline
         Interval & p0 & p1 & p2 & p3 & p4 \\
         \hline
         \smallskip
         Parameter & \\
         & \multicolumn{5}{c}{\em hot} \\
         \smallskip
         N$_H^{cold}$ (10$^{22}$ cm$^{-2}$) & 6.0$\pm$0.1 & 6.3$\pm$0.2 & 6.4$\pm$0.1 & 5.9$\pm$0.1 & 6.0$\pm$0.1\\
         \smallskip
         & \multicolumn{5}{c}{\em blackbody} \\
         \smallskip
         kT$_{bb}$ (keV) & 1.23$\pm$0.01 & 1.11$\pm$0.02 & 1.18$\pm$0.01 & 1.18$\pm$0.01 & 1.19$\pm$0.01 \\
         Norm (10$^{13}$ cm$^2$) & 0.97$\pm$0.04 & 0.2\,$\pm$\,0.1 & 1.00$\pm$0.05 & 0.73$\pm$0.05 & 0.88$\pm$0.06 \\
         \smallskip
         & \multicolumn{5}{c}{\em comt} \\
         \smallskip
         kT$_{bb}$ (keV) & (c) & (c) & (c) & (c) & (c) \\
         kT$_{hot}$ (keV)& 2.92$\pm$0.04 & 2.92$\pm$0.02 & 2.91$\pm$0.03 & 2.98$\pm$0.03 & 2.92$\pm$0.04 \\
         $\tau$ & 10 (f) & 10 (f) & 10 (f) & 10 (f) & 10 (f)  \\
         Norm (10$^{44}$ ph~s$^{-1}$~keV$^{-1}$)& 22$\pm$2 & 40$\pm$3 & 27$\pm$2 & 29$\pm$2 & 27$\pm$2 \\
         \smallskip
         & \multicolumn{5}{c}{\em pion$_1$} \\
         \smallskip
         N$_H^{pion}$ (10$^{22}$ cm$^{-2}$) & 9$\pm$1 & 9$\pm$4 & 16$\pm$5 & 13$\pm$4 & 16$\pm$2 \\
         log $\xi$ (erg~cm~s$^{-1}$) & 3.58$\pm$0.05 & 3.99$\pm$0.08 & 4.02$\pm$0.07 & 3.96$\pm$0.06 & 3.74$\pm$0.05 \\
         $\sigma_v$ (km~s$^{-1}$) & 68$\pm$8 & 49$\pm$14& 60$\pm$12 & 103$\pm$17 & 141$\pm$7 \\
         v (km~s$^{-1}$) & -194$\pm$8 & -393$\pm$12 & -459$\pm$11 & -393$\pm$15 & -218$\pm$9 \\
         Solid angle & 0.12$\pm$0.02 & 0.09$\pm$0.06 & 0.17$\pm$0.04 & 0.22$\pm$0.04 & 0.15$\pm$0.01 \\
         \hline \\
         Lumin$_{0.013-13.6~keV}^{total}$ (erg~s$^{-1}$) & 5.5$\times$10$^{37}$ & 6.1$\times$10$^{37}$ & 5.8$\times$10$^{37}$ & 5.9$\times$10$^{37}$ & 5.7$\times$10$^{37}$ \\
         \hline \\
         $Cstat$ (d.o.f.) & 3268 (2924) & 2946 (2726) & 3436 (3021) & 2997 (2840) & 3003 (2861) \\
         $\Delta Cstat$ (d.o.f.) (compared to Model 2a) & -13 (1) & 0 (1) & -10 (1) & -25 (1) & -108 (1) \\
         \hline\\
        Cross-normalization NuSTAR FPMA & 0.98 & 1.01 & 1.00/1.08 & 1.01 & 1.00 \\
        Cross-normalization NuSTAR FPMB & 0.98 & 1.00 & 1.03/1.02 & 1.00 & 0.99 \\
        \hline\\
    \end{tabular}
    \label{tab:specfits1c}
\end{table*}

Looking at interval p4, the improvement of models 2b and 2c with respect to 2a is due to the former models being able to capture the high column density of the absorber. As indicated in previous section, the column density for \fetfive\ and \fetsix\ is the highest for this interval based on fits to the K$\beta$ transitions. The K$\alpha$ lines are saturated, but Model 2a is not able to account for that because the lines are not reaching zero intensity likely due to scattered emission from the wind\footnote{We remark that emission from scattering in the wind is local to the source. This component is different to the previously observed dust scattered emission (see Sect.~\ref{sec:src} and \ref{sec:spectral}), which arises through scattering of the overall source emission in dust clouds in the interstellar medium along the line of sight.} setting a ``floor'' for the lines higher than zero. In contrast, both models 2b and 2c ``set such a floor'', with Model 2b doing it via the ``uncovered'' emission and Model Model 2c via plasma re-emission. Interestingly, these two models give similar improvements in $C-stat$. Model 2c is able to better account for the stronger Ly$\alpha$2 line compared to Ly$\alpha$1 in p2 but predicts also a stronger Ly$\alpha$2 line in p0 and p3, in which it is not observed. Model 2b provides a better overall fit to p4, although the structure (double-peak) of the Ly$\alpha$1 transition is better accounted for in Model 2c. 

Despite the described limitations, in general, Models 2b and 2c provide a good description of the data. Allowing both the covering factor and the solid angle to vary at the same time only introduces improvements of the fit of $\Delta\,C-stat$~=~2--3 for one additional d.o.f. We also attempted to verify Model~2b by substituting the covering factor (that is leaving it to one) by an independently added diffuse scattered component ({\tt hot*pion*(bb+comt)+hot(bb+comt)}, Model 3). This model provides fits of similar quality to Model 2b (with $\Delta\,C-stat$\,=\,0-9 for one additional d.o.f.). We also tried to add to Model~3 a plasma re-emission component with parameters independent from the absorbing plasma, to account for the averaging of plasma emission from other parts of the disc but we only obtain a fit improvement for intervals p1 and p3, indicating that we may have reached some degeneracy in the fits.
Overall, we observe scattering fractions of $\sim$20-30\%, which match well the amount of ``uncovered'' emission in Model~2b of 17-35\% and the solid angle of $\sim$10-20\% in Model~2c, thus indicating that with the current data we cannot distinguish further between alternative models.
 A potential exception is interval p0, for which a blue wing in the Ly$\alpha$1 transition is not accounted for by Models~2b-2c (see below).

During p1--p3, the ionisation and column density of the absorber are consistent among the three intervals within the errors, and the velocity width is generally small, $\sim$50 -- 110~km~$^{-1}$, and shows a trend of increasing width from p1 to p3. The most remarkable  change among these intervals is the blueshift of the absorber, which changes as a function of phase, indicating orbital modulation and reproducing the behaviour already observed in the fits to the individual lines in previous section.

For intervals p0 and p4, as we leave and approach the dipping phase, respectively, the ionisation of the observer significantly decreases compared to p1--p3. This decrease of ionisation may account for the increase of column density in \fetfive\ and \fetsix\ observed in Table~\ref{tab:slabspecfits}. For p0 the shift of the absorber is also remarkably larger, by $\sim$45~km~s$^{-1}$, for models 2b-2c compared to that found with fits to the individual lines. This and the lower ionisation of the plasma for these intervals may be pointing out to the presence of additional absorbers. 

\begin{figure*}
\centering
\includegraphics[width=1.0\textwidth,angle=0]{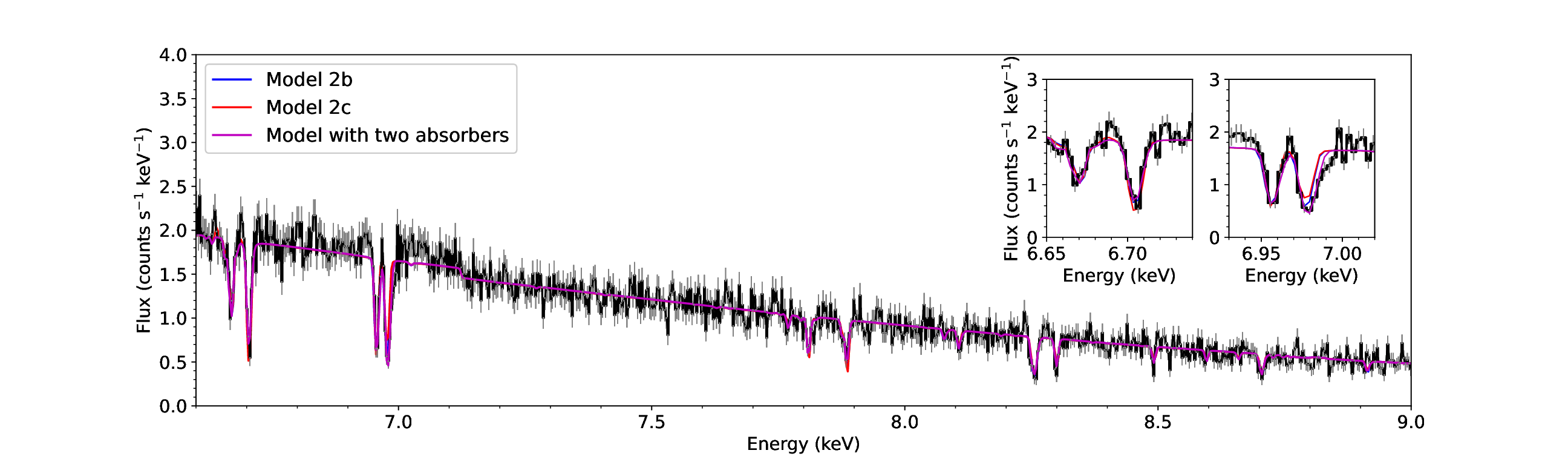}
\caption{Spectrum for interval p0 with the model with two absorbers displayed in Table~\ref{tab:specfits2} overplotted in magenta for comparison with models 2b and 2c with one absorber overplotted in blue and red, respectively (see Tables~\ref{tab:specfits2}, \ref{tab:specfits1b}--\ref{tab:specfits1c} for fitted parameters). The insets show a detail of the \fetfive\ and \fetsix\ K$\alpha$ regions highlighting the different prediction of the three models for the \fetsix\ spin-orbit doublet ratio. 
}
\label{fig:fcov_vs_omega3}
\end{figure*}

Therefore, we next attempted to add a second absorber for p0 (for which the differences in log~$\xi$ and $v$ are larger) to test the hypothesis that the lower ionisation of the plasma is actually caused by a mix of two plasmas, one highly ionised and similar to that found in p1--p3 and another one with lower ionisation resulting for example from some material from the dipping phase spreading outside of such phase. To constrain the fit we fix log\,$\xi$~=~4~erg~cm~s$^{-1}$ and $\sigma_v$ = 50 km~s$^{-1}$ for the first absorber, under the assumption that such ionised absorber does not significantly change among intervals. Adding a second absorber significantly improves the quality of the fit ($\Delta C-stat$ = 38 and 68 for 3 d.o.f. with respect to models 2b and 2c, respectively) by partially fitting the blue wing of the Ly$\alpha$1 line (see Table~\ref{tab:specfits2} and Fig.~\ref{fig:fcov_vs_omega3}). The two absorbers have now velocities of $\sim$-310 and -140~km~s$^{-1}$, thus the low ionisation absorber adopts a velocity closer to that obtained with the fit to the K$\beta$ lines, although we suspect that at this point we are reaching some degeneracy in the fits. 
We also attempted to add a second absorber for p4 but in this case the fit does not significantly improve with respect to the fit with one absorber.

\begin{table}
    \centering
        \caption{Spectral fits of the p0 interval with the {\tt hot*pion$_2$*pion$_1$*(bb+comt)} model. For each interval we fit the 2--10~keV XRISM and 4--30~keV NuSTAR spectra simultaneously. $f$ indicates a fixed parameter and $c$ a coupled parameter. The cross-normalization for NuSTAR with respect to Resolve is indicated in the last rows.}

    \begin{tabular}{lc}
        \hline
         Interval & p0 \\
         \hline
         \smallskip
         Parameter & \\
         & {\em hot} \\
         \smallskip
         N$_H^{cold}$ (10$^{22}$ cm$^{-2}$) & 6.0\,$\pm$\,0.1 \\
         \smallskip
         & {\em blackbody} \\
         \smallskip
         kT$_{bb}$ (keV) & 1.23\,$\pm$\,0.01 \\
         Norm (10$^{13}$ cm$^2$) & 1.0\,$\pm$\,0.5 \\
         \smallskip
         & {\em comt} \\
         \smallskip
         kT$_{bb}$ (keV) & (c) \\
         kT$_{hot}$ (keV)& 2.91\,$\pm$\,0.04 \\
         $\tau$ & 10 (f) \\
         Norm (10$^{44}$ ph~s$^{-1}$~keV$^{-1}$) & 24\,$\pm$\,2 \\
         \smallskip
         & {\em pion$_1$} \\
         \smallskip
         N$_H^{pion}$ (10$^{22}$ cm$^{-2}$) & 5\,$\pm$\,2 \\
         log $\xi$ (erg~cm~$s^{-1}$) & 4 (f) \\
         $\sigma_v$ (km~$s^{-1}$) & 50 (f) \\
         v (km~$s^{-1}$) & -307\,$\pm$\,44 \\
         Solid angle & 0.16\,$\pm$\,0.07 \\
         \smallskip\\
         & {\em pion$_2$} \\
         \smallskip
         N$_H^{pion2}$ (10$^{22}$ cm$^{-2}$) & 17\,$\pm$\,4 \\
         log~$\xi$ (erg~cm~$s^{-1}$) & 3.51\,$\pm$\,0.06 \\
         $\sigma_v$ (km~$s^{-1}$) & 51\,$\pm$\,10 \\
         v (km~$s^{-1}$) & -139\,$\pm$\,15  \\
         Covering fraction & 0.60\,$\pm$\,0.09 \\
         \hline \\
         Lumin$_{0.013-13.6~keV}^{total}$ & 5.8$\times$10$^{37}$ \\
         \hline \\
         C-stat (d.o.f.) & 3200 (2921) \\
         \hline\\
        Cross-normalization NuSTAR FPMA & 0.98 \\
        Cross-normalization NuSTAR FPMB & 0.98 \\
        \hline\\
    \end{tabular}
    \label{tab:specfits2}
\end{table}  

\section{Discussion}

The unprecedented resolution of the microcalorimeter Resolve onboard \xrism\ has enabled to determine for the first time a significant blueshift in the highly photoionised absorber previously seen in \src, indicating the presence of an outflow with a column density of $\sim$1-2$\times$10$^{23}$~cm$^{-2}$. 

Phase-resolved analysis shows that the highly ionised plasma is present in all intervals of the persistent emission. The plasma consistently shows a blueshift in all intervals, and the value is modulated by the orbital phase. The velocity width is generally low, down to values of 50~km~s$^{-1}$ in p1, indicating minimal sheer and/or turbulence, although it increases in p4 up to $\sim$150~km~s$^{-1}$, perhaps associated to an increase of turbulence as we approach the dipping phase. Solid angles of 10-20\% are found with Model~2c, consistent with an equatorial wind, and with estimates from the solid angle of the wind in the LMXB GX~13+1 \citep{gx13:xrism}. With Model~2b, we find covering factors of $\sim$65-83\%, which are smaller for p2 and p3 compared to p0 and p4, that is, the ``uncovered'' (or scattered) emission is larger for p2 and p3. If we fit only the \xrism/Resolve data with an additional scattered component,  
we find scattering fractions of 20-30\%, which are again larger for p2 and p3 compared to p0 and p4. We interpret this as p2 and p3 (at orbital phases 0.30-0.45 and 0.45-0.60, respectively) seeing more  diffuse scattered emission from the opposite part of the disc (phases 0.80-0.95 and 0.95-1.10) due to the opacity of additional material present during the dipping phase, especially affecting p2. However, we emphasize that in reality we expect both scattered continuum and plasma re-emission (including that from plasma in other parts of the disc). Therefore, the overall scattering fraction is likely to be slightly less than found when only one of the components (continuum scattering or plasma re-emission) is included.  
Interval p1 is an outlier with respect to such trends but it also shows continuum parameters deviating from those seen during the other intervals. We attribute the outlier behaviour of p1 to the worse statistics of that interval because of observation~2 largely missing such phase, or to a higher short-term flux during observation~1 (see Fig.~\ref{lc}). Moreover, fits with the continuum tied among intervals p1-p3 result in similar values for the absorbers to those observed in Models 2b-2c. Therefore, overall, changes in the scattering and absorption in the line of sight, rather than source intrinsic variability, are likely to fully explain the variability across the orbital phase and naturally explain why the light curves remain stable over periods of days. 

In interval p0, a second, lower ionisation, plasma appears with a lower outflow velocity than the highly ionised plasma. Since in p0 we are leaving the dipping phase, such a plasma may be the less ionised component typically seen during dips \citep{ionabs:diaz06aa}, which is spreading over to p0 (see Caruso et al. (in prep) for the detailed analysis of the dips). Introducing this second absorber allows the highly ionised plasma to better account for the blue wing of the \fetsix\ Ly$\alpha$1 transition (although not completely). We note that blue wings associated to the \fetsix\ Ly$\alpha$1 transition are apparent in other XRISM observations of LMXBs \citep[see Fig.~6 of ][]{tsujimoto25pasj}. The phase-resolved analysis performed here indicates that despite blue wings being predicted by magnetic wind models \citep[e.g. ][]{Fukumura2017}, such association is unlikely in this case, since the wing should be seen across all phase-resolved spectra. Instead, since blue wings are more apparent in observations in which high plasma opacities are found \citep{gx13:xrism}, we suggest that our relatively simple models assuming optically thin, slab-like, plasmas are not able to account in detail for the complexity of radiation transfer and scattering effects at high opacities. We emphasize that sources such as \src\ are ideal to understand these effects due to their stable luminosity and accretion state affording phase-resolved studies that isolate the phenomenology due to the orbital dependence of the photoionised plasmas.

\subsection{Radial velocity}
\label{rv}

\begin{figure*}
\centering
\includegraphics[width=0.48\textwidth]{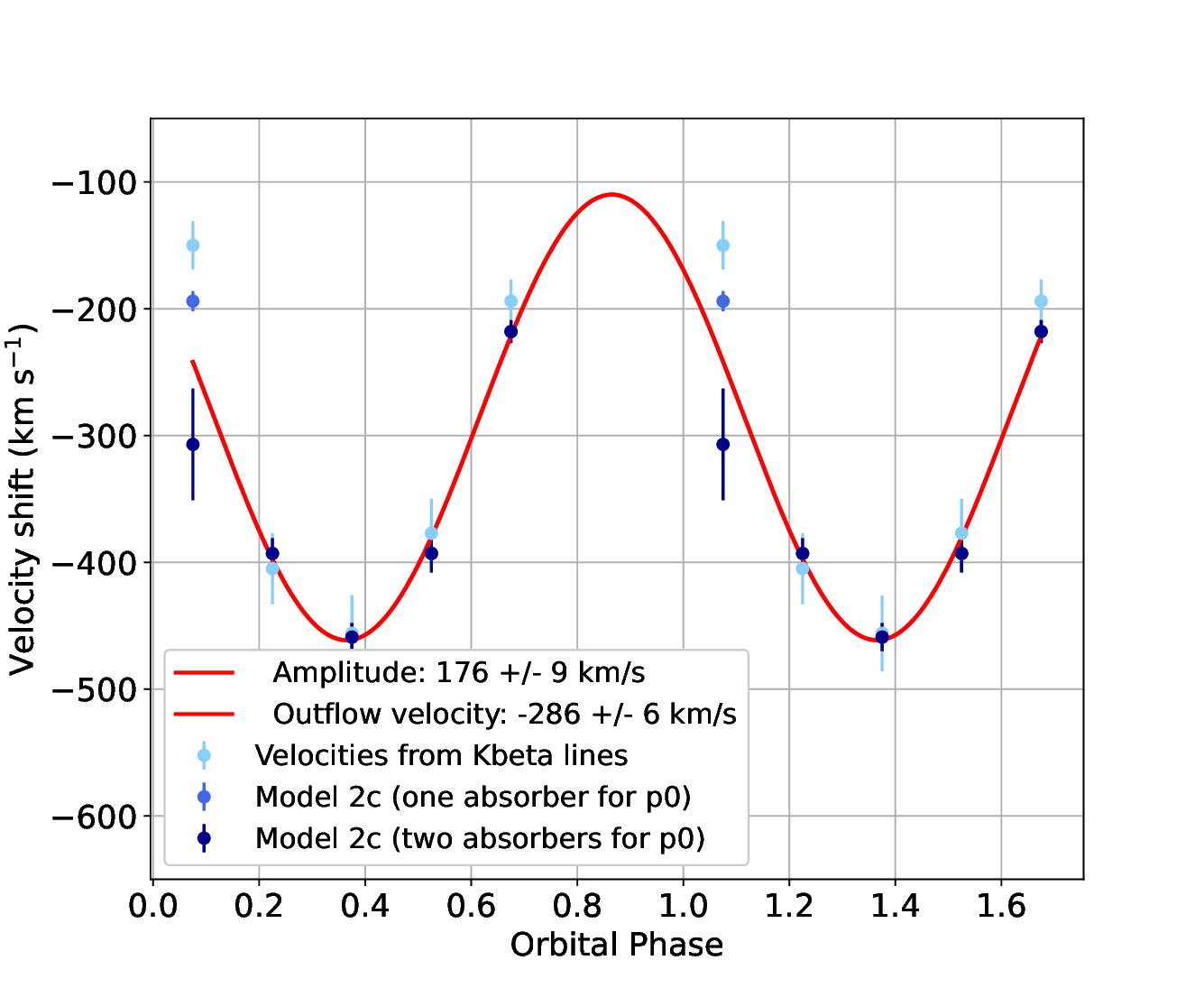}
\includegraphics[width=0.49\textwidth]{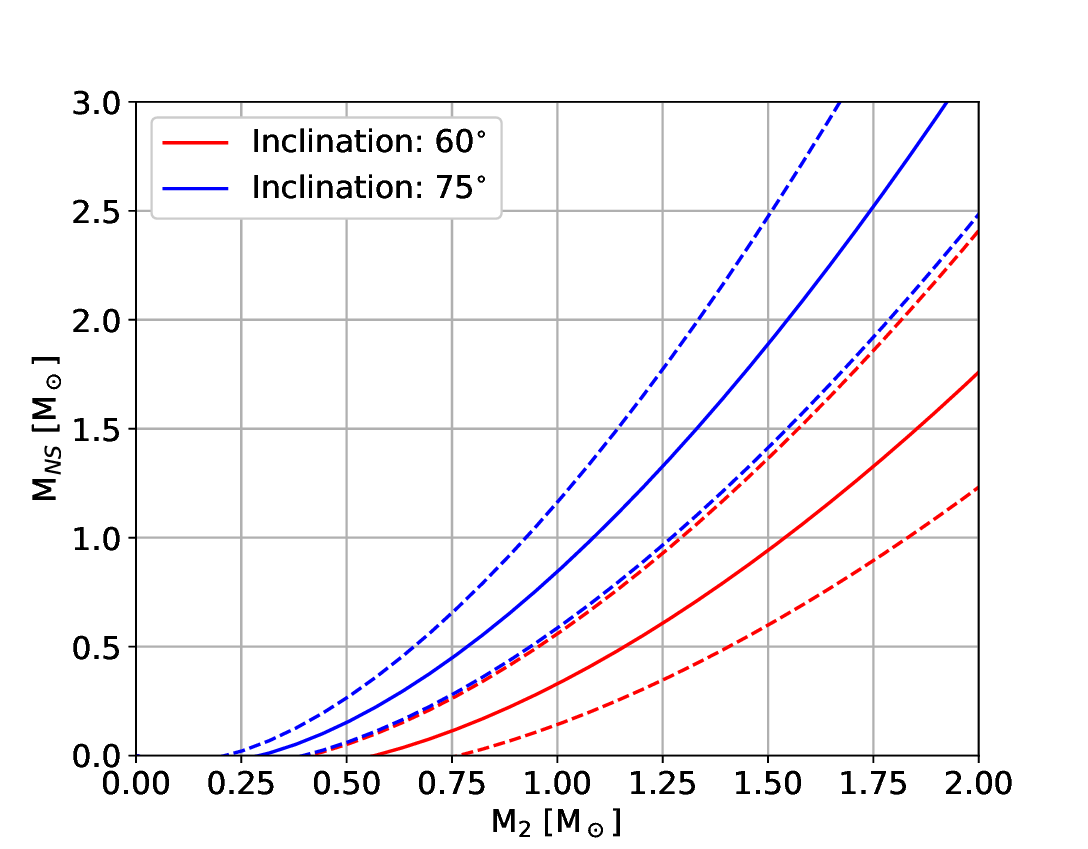}
\caption{(Left) Best-fit values of the velocity shift of the highly ionised absorber as a function of orbital phase as determined with three different methods (see text) and best-fit model (red) using the values from Model 2c with two absorbers for p0 (dark blue). (Right): Allowed masses for the companion star for two different inclinations based on the best-fit model from the RV curve.}
\label{fig:radialv}
\end{figure*}

The radial velocity (RV) dependence on orbital phase shows a clear sinusoidal pattern (see Fig.~\ref{fig:radialv}). However, the values of the velocity for interval p0 significantly differ for the different models. We measure -150\,$\pm$\,19~km~s$^{-1}$ when using the \fetfive\ and \fetsix\ K$\beta$ transitions (Table~\ref{tab:slabspecfits}) and -194\,$\pm$\,8~km~s$^{-1}$ when using Model 2c (Table~\ref{tab:specfits1c}). However, the largest difference is found for the fit with two pion components, in which the highly ionised component (with ionisation and velocity width similar to the p1-p3 intervals) results in an outflow velocity of -307\,$\pm$\,44~km~s$^{-1}$ (Table~\ref{tab:specfits2}). Fitting the data with a sine curve thus results in different values for the amplitude and the offset. We obtain RV amplitudes of 229$\pm$14\,km~s$^{-1}$, 200$\pm$6\,km~s$^{-1}$ and 177$\pm$9\,km~s$^{-1}$ and RV offsets of -245\,$\pm$\,8\,km~s$^{-1}$, -270\,$\pm$\,4\,km~s$^{-1}$ and -285\,$\pm$\,7\,km~s$^{-1}$ for the three outlined methods, respectively.

The RV offset is large, regardless of the method and even after considering the sum of the gain uncertainty and \xrism\ motion around the Earth (averaging to 0~km~s$^{-1}$), the Earth's motion around the Sun (23.8 and 23.0~km~s$^{-1}$ during the first/second observations, respectively) and the Sun's motion with respect to the Local Standard of Rest (2.8\,km~s$^{-1}$). 
Since the companion of \src\ is not well characterised due to the high extinction, we use instead that \src\ follows the Galactic rotation except for the natal kick and that the Galactic rotation curve is flat at 220 km s$^{-1}$ \citep[e.g. ][]{mroz2019ApJ} to determine the estimated velocity. At a distance of $\sim$15~kpc \citep{1624:xiang07apj}, the relative velocity of \src\ is then -3~km s$^{-1}$. Assuming a natal kick of 54~km~s$^{-1}$ \citep{Hobbs2005} and accounting for all motions, we obtain a velocity of the absorber of $\sim$200-320~km ~s$^{-1}$, thus firmly establishing the presence of a highly ionised outflow in \src.

We next use the RV amplitude to set some constraints on the mass of the companion. We determine the mass function as f~=~K$^3$~P$_{orb}$/2$\pi$G, where K is the semi-amplitude of the RV curve and G is the gravitational constant. This results in values of f of 1.08, 0.71 and 0.50 M$_\odot$ for the three methods. We assume an inclination between 60$^{\circ}$ and 75$^{\circ}$ from the presence of absorption dips and the absence of eclipses \citep{frank87aa}. Using the relation f = M$_{2}^{3}$ sin$^{3}$($i$) / M$_{tot}^2$, we can then determine the mass of the companion as a function of the neutron star mass. For plausible values of the neutron star mass of 1.3-1.5~M$_\odot$, we obtain a mass for the companion star near 1~M$_\odot$ only when two absorbers are considered for p0, namely a mass of the companion of 1.3-1.4~M$_\odot$ for an inclination of 75$^{\circ}$. In contrast, for a mass function of 0.71 the corresponding companion star masses are $>$1.8~M$_\odot$. Therefore, the scenario in which a second intervening absorber with lower velocity appears in the line of sight between the wind and us is more supportive of \src\ being a low-mass X-ray binary, although an intermediate X-ray binary could be a possibility for one absorber.

We note that the while K corresponds to the RV semiamplitude of the neutron star, in reality, we are measuring absorption in a wind launched far from the neutron star. However, we estimate that this does not introduce systematics in the measurement due to the fact that we are using absorption lines, and that the wind is most likely axisymmetric and centred around the neutron star. While a precessing disc could still introduce some asymmetry, such a disc has not been claimed so far.

Finally, we also note that the zero orbital phase is offset by $\sim$0.1 in the RV curve, indicating that with our definition of zero phase (see Sect.~\ref{sec:phase-resolved}), p0 would be almost centred around eclipse. Methods like Doppler tomography may help us clarify further the emission zones.  

\subsection{Origin of the wind}

Based on the orbital modulation and the RV analysis performed in Sect.~\ref{rv}, we estimate the velocity of the outflow to be $\sim$200--320~km s$^{-1}$. 

The cleanest characterisation of the outflow can be done at the orbital phase farthest from the dip, that is, during interval p2, due to the absence of other intervening absorbers. Therefore, we use the source luminosity, and column density and ionisation of the absorber obtained during that interval to estimate the launching radius of the outflow. 

The luminosity of \src\ reaches $\sim$6$\times$10$^{37}$~erg~s$^{-1}$ in these observations, corresponding to 0.32 L$_{Edd}$ for a typical neutron star mass of 1.4~M$\odot$. 
Using $log~\xi$~=~$L n^{-1}r^{-2}$, where $n$ and $r$ are the density and the launching radius of the wind, respectively, and considering $\Delta r/r~\leq$\,1, we obtain that the wind is launched at a distance of $\sim$\,3.9$\times$10$^{10}$~cm. Considering a Compton radius of 10$^{11}$ cm for a Compton temperature of $\sim$10$^7$~K \citep[see eq. 2.7 in ][]{begelman83apj}, it follows that the wind is being launched at $\sim$0.39 of the Compton radius, in the region allowed for Compton heated winds. This is only slightly larger than the location estimated for the origin of the \fetsix\ absorption lines of 3$\times$10$^{10}$~cm based on previous phase-resolved analysis with Chandra/HETGS, for which no significant line shifts were observed \citep{1624:xiang09apj}.

At such a large radius, the gravitational and transverse Doppler shift are expected to be negligible. Thus, the observed velocity is expected to correspond to the escape velocity of the wind. However, at a radius of 3.9$\times$10$^{10}$ cm, the escape velocity, estimated as v$_{esc}$ = $\sqrt{2*G*M/r}$ where $G$ is the gravitational constant, $M$ is the mass of the neutron star and $r$ is the estimated wind launching radius, is $\sim$\,980~km\,s$^{-1}$, more than three times the outflow estimated velocity. This could indicate projection effects, albeit implying seeing the wind at a large offset angle, or some deceleration of the wind after being launched. A third possibility is that the bulk of the ion column densities for the relevant ions is at a radius larger than the launching radius. In that case, the blueshift of the lines would be dominated by those at the radius where the column density peaks \citep[see ][Fig.~6]{tomaru20mnras}.

Remarkably, the width of the lines is only $\sim$50-70~km s$^{-1}$ during p0-p2. This indicates that no stratification is present, thus supporting the origin of the wind as thermal pressure with the bulk of the wind located over a small range of radii. Moreover, the lines are narrow also when compared to other X-ray binaries already observed with \xrism\ \citep{tsujimoto25pasj}. 
However, the changes observed throughout the orbit emphasize the need of phase-resolved spectroscopy to precisely determine the line widths. Not only do spectra aggregated throughout the orbit artificially broaden the lines due to the sum of different line shifts but we have also demonstrated that at phases close to the dipping events additional absorbers and/or turbulence increase the line widths, which could be wrongly interpreted if their origin is uniquely attributed to the outflow. In addition, a larger width could arise due to saturation of the lines derived from large column densities if not properly accounted for.

We note that additional absorbers may not uniquely be related to dipping events locked in phase. For example, the light curve during intervals p0 and p4 does not show any strong absorption events reminiscent of dips. However, the spectroscopic analysis already reveals an increase in the column density of the absorber derived from a decrease of its ionisation probably due to mixing with material from the accretion stream or the bulge. If additional absorbers are behind the increase of the line width in p4, this could also explain the larger widths observed for the black hole 4U~1630$-$47 \citep{miller25apj} for which an additional absorber was also seen in part of the orbit. 

We finally estimate the mass outflow rate making use of the precise knowledge of luminosity (afforded by the accurate distance and the broadband spectral energy distribution), and the measured ionisation parameter and solid angle. Again, we use interval p2 as having the most reliable parameters and obtain a mass outflow rate of $\dot M \sim n~m_p~r^2~v_{out}$ = 5--10\,$\times$\,10$^{16}$\,g~s$^{-1}$ (with the higher estimate taking into account the outflow from both sides of the disc). 
This mass outflow rate is comparable to that estimated for GX~13+1 at a similar Eddington luminosity \citep{gx13:ueda04apj}.

\subsection{Comparison to other systems}

The \xrism\ PV phase included observations of two other high inclination LMXBs for which winds had been already reported, GX~13+1 \citep[e.g.][]{gx13:ueda04apj} and Cir~X-1 \citep[e.g.][]{cirx1:brandt00apjl}. 

For GX~13+1, previous observations at 0.3-0.4 L$_{Edd}$ had repeatedly shown a wind \citep[][]{gx13:ueda04apj,gx13:allen18apj} with a relatively high, changing, column density \citep{gx13:diaz12aa}. The \xrism\ observation, during which the source had a luminosity near Eddington, revealed an unprecedented increase in the column density, making the wind optically thick and creating significant obscuration \citep{gx13:xrism}. Interestingly, the wind observed in \src\ is similar to that from previous observations of GX~13+1, indicating that at similar luminosities, winds are alike in terms of opacity, ionisation and outflow velocity. 

Cir~X-1 also showed a luminosity of $\sim$0.4~L$_{Edd}$ during the XRISM PV phase observation, and again, the outflow velocity was $\sim$\,300~km~s$^{-1}$, although higher velocities had been observed in previous Chandra observations at higher luminosities \citep{cirx1:brandt00apjl, cirx1:schulz02apj}.

Thus, remarkably, the outflow velocities observed for GX~13+1, Cir~X-1 and \src\ are very similar for luminosities of 0.3-0.4~L$_{Edd}$, despite the large differences in the orbital period of these systems, ranging from 24 days to less than one day. Small differences could arise due to velocity projections for slightly different inclinations or different systemic velocities. However, the general trend points to these winds being launched via the same thermal-radiative pressure mechanism, for which the launching radius only depends on the spectral hardness and the source luminosity. 

Similarly to what was observed by Chandra for Cir~X-1, higher luminosities would result in higher outflow velocities \citep{cirx1:brandt00apjl, cirx1:schulz08apj}, and therefore, the higher luminosity of GX~13+1 in the \xrism\ observation would result in the appearance of a faster, broader, outflow component \citep{gx13:xrism}. However, we emphasize that blue wings are not necessarily associated to fast components, since for \src\ a blue wing for the \fetsix\ Ly$\alpha$1 component of the spin-orbit doublet is phase-dependent. If such wing were due to the structure of the wind and assuming that the wind should be axisymmetric, we would expect to see it also at other phases. An alternative origin for the blue wing could be the superposition of lower ionization, high column density, material from the accretion disc rim and the accretion disc wind. In intervals approaching the absorption dip, the mixing of such components can result in different profiles depending on the line saturation and the velocity superposition of the two components. An increase of turbulence as we approach the dip could also explain the increase of line widths.

A strong scattering component is common to all the observed systems. This results in absorption lines not reaching zero intensity despite the large column densities. Including a scattered component in broadband spectra is however difficult due to the energy dependence of Compton scattering not being taken into account in a model that simply adds a fraction of the initial continuum. For \src, we find that allowing for a covering factor of the wind less than one improves significantly the fits and we attribute it to the ``uncovered'' emission mimicking the scattered component. Overall, scattering fractions between 5 and 30\% are reported, consistent with the opacity expected from the wind and additional material at the disc rim.

Besides the scattered continuum we expect that re-emission from the plasma reaches us from all the disc, and therefore that the re-emission should in principle correspond to relatively broad emission lines with lower velocity than the absorption lines. This seems to be the case for GX~13+1 \citep{gx13:xrism}, consistent with  interpretations based on previous low-resolution observations \citep{gx13:diaz12aa}. In contrast, for \src, we find that the changes in the ratio of the \fetsix\ spin-orbit doublet require some narrow line re-emission taking place. 

\subsection{Limitations}

The overall spectral analysis results in a relatively simple physical picture according to which a highly ionised thermal-radiative outflow is modulated by orbital motions and mixes with lower ionisation plasmas near the dipping phase, as shown for the particular case of p0. However, some additional plasma structure has likely not been captured with this simple model. Despite the excellent statistics of the observations presented here, the need to perform phase-resolved spectroscopy naturally reduces the overall statistics of each spectrum. Remaining small residuals indicate that we may be reaching the limit of these data to disentangle the highly ionised wind from the lower ionisation plasma expected from the accretion stream or material spreading from the bulge and which causes the absorption during dips. 

\begin{acknowledgements}
MDT thanks A. Gnarini and S. Bianchi for providing an accurate ephemeris for the dips of the IXPE observation and R. Tomaru for discussions on thermal winds.
TY acknowledges support by NASA under award number 80GSFC24M0006. We thank the referee for a constructive and insightful report.
\end{acknowledgements}

%
\bibliographystyle{aa} 
\bibliography{1624} 

@ARTICLE{1323:boirin05aa,
    author = {{Boirin}, L. and {M\'endez}, M. {D{\'i}az Trigo}, M. and {Parmar}, A. N. and {Kaastra}, J.},
    title = "{A highly-ionised absorber in the X-ray binary 4U\,1323$-$619: a
    new explanation of the dipping phenomena}",
    journal = {\aap},
    year = 2005,
    month = feb,
    volume = {436},
    pages={195},
    adsurl = {unknown}
}

@ARTICLE{1624:smale01apj,
   author = {{Smale}, A.~P. and {Church}, M.~J. and {Ba{\l}uci{\'n}ska-Church}, M.
    },
    title = "{The Ephemeris and Dipping Spectral Behavior of 4U 1624-49}",
  journal = {\apj},
   eprint = {astro-ph/0010397},
     year = 2001,
    month = apr,
   volume = 550,
    pages = {962-969},
      doi = {10.1086/319800},
   url = {http://adsabs.harvard.edu/cgi-bin/nph-bib_query?bibcode=2001ApJ...550..962S&db_key=AST},
}

@ARTICLE{1624:parmar02aa,
     author = {{Parmar}, A.~N. and {Oosterbroek}, T. and {Boirin}, L. and {Lumb}, D.
     },
     title = "{Discovery of narrow X-ray absorption features from the dipping low-mass X-ray binary X 1624-490 with XMM-Newton}",
     journal = {\aap},
     year = 2002,
     month = may,
     volume = 386,
     pages = {910-915},
     url = {http://cdsads.u-strasbg.fr/cgi-bin/nph-bib_query?bibcode=2002A%26A...386..910P&db_key=AST},
 }

@ARTICLE{1624:xiang07apj,
       author = {{Xiang}, Jingen and {Lee}, Julia C. and {Nowak}, Michael A.},
        title = "{Using the X-Ray Dust Scattering Halo of 4U 1624-490 to Determine Distance and Dust Distributions}",
      journal = {\apj},
         year = 2007,
        month = may,
       volume = {660},
       number = {2},
        pages = {1309-1318},
          doi = {10.1086/513308},
archivePrefix = {arXiv},
       eprint = {astro-ph/0701865},
 primaryClass = {astro-ph},
       adsurl = {https://ui.adsabs.harvard.edu/abs/2007ApJ...660.1309X}
}

@ARTICLE{1624:xiang09apj,
   author = {{Xiang}, J. and {Lee}, J.~C. and {Nowak}, M.~A. and {Wilms}, J. and 
	{Schulz}, N.~S.},
    title = "{The Accretion Disk Corona and Disk Atmosphere of 4U 1624-490  as Viewed by the Chandra-High Energy Transmission Grating Spectrometer}",
  journal = {\apj},
archivePrefix = "arXiv",
   eprint = {0905.3925},
 primaryClass = "astro-ph.HE",
     year = 2009,
    month = aug,
   volume = 701,
    pages = {984-993},
      doi = {10.1088/0004-637X/701/2/984},
   adsurl = {http://adsabs.harvard.edu/abs/2009ApJ...701..984X}
}

@ARTICLE{gx13:ueda04apj,
   author = {{Ueda}, Y. and {Murakami}, H. and {Yamaoka}, K. and {Dotani}, T. and
    {Ebisawa}, K.},
    title = "{Chandra High-Resolution Spectroscopy of the Absorption-Line Features in the Low-Mass X-Ray Binary GX 13+1}",
  journal = {\apj},
     year = 2004,
    month = jul,
   volume = 609,
    pages = {325-334},
   adsurl = {http://adsabs.harvard.edu/cgi-bin/nph-bib_query?bibcode=2004ApJ...609..325U&db_key=AST}
}

@ARTICLE{gx13:diaz12aa,
   author = {{D{\'{\i}}az Trigo}, M. and {Sidoli}, L. and {Boirin}, L. and 
	{Parmar}, A.~N.},
    title = "{XMM-Newton observations of GX 13 + 1: correlation between photoionised absorption and broad line emission}",
  journal = {\aap},
archivePrefix = "arXiv",
   eprint = {1204.5904},
 primaryClass = "astro-ph.HE",
     year = 2012,
    month = jul,
   volume = 543,
      eid = {A50},
    pages = {A50},
      doi = {10.1051/0004-6361/201219049},
   adsurl = {http://adsabs.harvard.edu/abs/2012A%26A...543A..50D}
}

@ARTICLE{gx13:allen18apj,
   author = {{Allen}, J.~L. and {Schulz}, N.~S. and {Homan}, J. and {Neilsen}, J. and 
	{Nowak}, M.~A. and {Chakrabarty}, D.},
    title = "{The Disk Wind in the Neutron Star Low-mass X-Ray Binary GX 13+1}",
  journal = {\apj},
archivePrefix = "arXiv",
   eprint = {1806.08800},
 primaryClass = "astro-ph.HE",
     year = 2018,
    month = jul,
   volume = 861,
      eid = {26},
    pages = {26},
      doi = {10.3847/1538-4357/aac2d1},
   adsurl = {http://adsabs.harvard.edu/abs/2018ApJ...861...26A}
}

@ARTICLE{cirx1:brandt00apjl,
    author = {{Brandt}, W.~N. and {Schulz}, N.~S.},
    title = "{The Discovery of Broad P Cygni X-Ray Lines from Circinus X-1 with the Chandra High-Energy Transmission Grating Spectrometer}",
    journal = {\apjl},
    year = 2000,
    month = dec,
    volume = 544,
    pages = {L123-L127},
    adsurl = {http://cdsads.u-strasbg.fr/cgi-bin/nph-bib_query?bibcode=2000ApJ...544L.123B&db_key=AST},
}

@ARTICLE{cirx1:schulz02apj,
    author = {{Schulz}, N.~S. and {Brandt}, W.~N.},
    title = "{Variability of the X-Ray P Cygni Line Profiles from Circinus X-1 near Zero Phase}",
    journal = {\apj},
    year = 2002,
    month = jun,
    volume = 572,
    pages = {971-983},
    adsurl = {http://cdsads.u-strasbg.fr/cgi-bin/nph-bib_query?bibcode=2002ApJ...572..971S&db_key=AST},
}

@ARTICLE{cirx1:schulz08apj,
   author = {{Schulz}, N.~S. and {Kallman}, T.~E. and {Galloway}, D.~K. and 
	{Brandt}, W.~N.},
    title = "{The Variable Warm Absorber in Circinus X-1}",
  journal = {\apj},
archivePrefix = "arXiv",
   eprint = {0709.3336},
     year = 2008,
    month = jan,
   volume = 672,
    pages = {1091-1102},
      doi = {10.1086/523809},
   adsurl = {http://adsabs.harvard.edu/abs/2008ApJ...672.1091S}
}

@ARTICLE{1655:miller06nat,
   author = {{Miller}, J.~M. and {Raymond}, J. and {Fabian}, A. and {Steeghs}, D. and 
	{Homan}, J. and {Reynolds}, C. and {van der Klis}, M. and {Wijnands}, R.
	},
    title = "{The magnetic nature of disk accretion onto black holes}",
  journal = {\nat},
   eprint = {arXiv:astro-ph/0605390},
     year = 2006,
    month = jun,
   volume = 441,
    pages = {953-955},
      doi = {10.1038/nature04912},
   adsurl = {http://adsabs.harvard.edu/abs/2006Natur.441..953M}
}

@ARTICLE{begelman83apj,
   author = {{Begelman}, M.~C. and {McKee}, C.~F. and {Shields}, G.~A.},
    title = "{Compton heated winds and coronae above accretion disks. I Dynamics}",
  journal = {\apj},
     year = 1983,
    month = aug,
   volume = 271,
    pages = {70-88},
      doi = {10.1086/161178},
   adsurl = {http://adsabs.harvard.edu/abs/1983ApJ...271...70B}
}

@ARTICLE{frank87aa,
    author = {{Frank}, J. and {King}, A.\ R. and {Lasota}, J.\ P.},
    title = {The light curves of low-mass X-ray binaries},
    journal = {\aap},
    year = 1987,
    month = may,
    volume = 178,
    pages = {137--142},
    url = {http://cdsads.u-strasbg.fr/cgi-bin/nph-bib_query?bibcode=1987A%26A...178..137F&db_key=AST},
}

@ARTICLE{woods96apj,
   author = {{Woods}, D.~T. and {Klein}, R.~I. and {Castor}, J.~I. and {McKee}, C.~F. and 
	{Bell}, J.~B.},
    title = "{X-Ray--heated Coronae and Winds from Accretion Disks: Time-dependent Two-dimensional Hydrodynamics with Adaptive Mesh Refinement}",
  journal = {\apj},
     year = 1996,
    month = apr,
   volume = 461,
    pages = {767},
      doi = {10.1086/177101},
   adsurl = {http://adsabs.harvard.edu/abs/1996ApJ...461..767W}
}

@ARTICLE{ionabs:diaz06aa,
    author = {{D{\'i}az Trigo}, M. and {Parmar}, A. N. and {Boirin}, L. and {M\'endez}, M. and
        {Kaastra}, J.},
    title = "{Spectral changed during dipping in low-mass X-ray binaries due to highly-ionized absorbers}",
    journal = {\aap},
    year = 2006,
    month = jan,
    volume = {445},
    pages = {179-189},
    url = {},
}

@ARTICLE{ponti12mnras,
   author = {{Ponti}, G. and {Fender}, R.~P. and {Begelman}, M.~C. and {Dunn}, R.~J.~H. and 
	{Neilsen}, J. and {Coriat}, M.},
    title = "{Ubiquitous equatorial accretion disc winds in black hole soft states}",
  journal = {\mnras},
archivePrefix = "arXiv",
   eprint = {1201.4172},
 primaryClass = "astro-ph.HE",
     year = 2012,
    month = may,
   volume = 422,
    pages = {L11},
      doi = {10.1111/j.1745-3933.2012.01224.x},
   adsurl = {http://adsabs.harvard.edu/abs/2012MNRAS.422L..11P}
}

@ARTICLE{diaz16an,
       author = {{D{\'\i}az Trigo}, M. and {Boirin}, L.},
        title = "{Accretion disc atmospheres and winds in low-mass X-ray binaries}",
      journal = {Astronomische Nachrichten},
         year = 2016,
        month = may,
       volume = {337},
       number = {4-5},
        pages = {368},
          doi = {10.1002/asna.201612315},
archivePrefix = {arXiv},
       eprint = {1510.03576},
 primaryClass = {astro-ph.HE},
       adsurl = {https://ui.adsabs.harvard.edu/abs/2016AN....337..368D}
}

@ARTICLE{Gnarini24,
       author = {{Gnarini}, Andrea and {Lynne Saade}, M. and {Ursini}, Francesco and {Bianchi}, Stefano and {Capitanio}, Fiamma and {Kaaret}, Philip and {Matt}, Giorgio and {Poutanen}, Juri and {Zhang}, Wenda},
        title = "{Constraining the geometry of the dipping atoll 4U 1624{\textendash}49 with X-ray spectroscopy and polarimetry}",
      journal = {\aap},
         year = 2024,
        month = oct,
       volume = {690},
          eid = {A230},
        pages = {A230},
          doi = {10.1051/0004-6361/202450716},
archivePrefix = {arXiv},
       eprint = {2408.02309},
 primaryClass = {astro-ph.HE},
       adsurl = {https://ui.adsabs.harvard.edu/abs/2024A&A...690A.230G}
}

@ARTICLE{Fukumura2017,
       author = {{Fukumura}, Keigo and {Kazanas}, Demosthenes and {Shrader}, Chris and {Behar}, Ehud and {Tombesi}, Francesco and {Contopoulos}, Ioannis},
        title = "{Magnetic origin of black hole winds across the mass scale}",
      journal = {Nature Astronomy},
         year = 2017,
        month = mar,
       volume = {1},
          eid = {0062},
        pages = {0062},
          doi = {10.1038/s41550-017-0062},
archivePrefix = {arXiv},
       eprint = {1702.02197},
 primaryClass = {astro-ph.HE},
       adsurl = {https://ui.adsabs.harvard.edu/abs/2017NatAs...1E..62F}
}

@ARTICLE{Neilsen2012,
       author = {{Neilsen}, Joseph and {Homan}, Jeroen},
        title = "{A Hybrid Magnetically/Thermally Driven Wind in the Black Hole GRO J1655-40?}",
      journal = {\apj},
         year = 2012,
        month = may,
       volume = {750},
       number = {1},
          eid = {27},
        pages = {27},
          doi = {10.1088/0004-637X/750/1/27},
archivePrefix = {arXiv},
       eprint = {1202.6053},
 primaryClass = {astro-ph.HE},
       adsurl = {https://ui.adsabs.harvard.edu/abs/2012ApJ...750...27N}
}

@INCOLLECTION{fender16lnp,
       author = {{Fender}, Rob and {Mu{\~n}oz-Darias}, Teo},
        title = "{The Balance of Power: Accretion and Feedback in Stellar Mass Black Holes}",
    booktitle = {Lecture Notes in Physics, Berlin Springer Verlag},
         year = 2016,
       editor = {{Haardt}, Francesco and {Gorini}, Vittorio and {Moschella}, Ugo and {Treves}, Aldo and {Colpi}, Monica},
       volume = {905},
        pages = {65},
          doi = {10.1007/978-3-319-19416-5_3},
       adsurl = {https://ui.adsabs.harvard.edu/abs/2016LNP...905...65F}
}

@ARTICLE{Kaastra2016,
       author = {{Kaastra}, J.~S. and {Bleeker}, J.~A.~M.},
        title = "{Optimal binning of X-ray spectra and response matrix design}",
      journal = {\aap},
         year = 2016,
        month = mar,
       volume = {587},
          eid = {A151},
        pages = {A151},
          doi = {10.1051/0004-6361/201527395},
archivePrefix = {arXiv},
       eprint = {1601.05309},
 primaryClass = {astro-ph.IM},
       adsurl = {https://ui.adsabs.harvard.edu/abs/2016A&A...587A.151K}
}

@ARTICLE{Hobbs2005,
       author = {{Hobbs}, G. and {Lorimer}, D.~R. and {Lyne}, A.~G. and {Kramer}, M.},
        title = "{A statistical study of 233 pulsar proper motions}",
      journal = {\mnras},
         year = 2005,
        month = jul,
       volume = {360},
       number = {3},
        pages = {974-992},
          doi = {10.1111/j.1365-2966.2005.09087.x},
archivePrefix = {arXiv},
       eprint = {astro-ph/0504584},
 primaryClass = {astro-ph},
       adsurl = {https://ui.adsabs.harvard.edu/abs/2005MNRAS.360..974H}
}

@ARTICLE{gx13:xrism,
       author = {{XRISM collaboration} and {Audard}, Marc and {Awaki}, Hisamitsu and {Ballhausen}, Ralf and {Bamba}, Aya and {Behar}, Ehud and {Boissay-Malaquin}, Rozenn and {Brenneman}, Laura and {Brown}, Gregory V. and {Corrales}, Lia and {Costantini}, Elisa and {Cumbee}, Renata and {Diaz Trigo}, Maria and {Done}, Chris and {Dotani}, Tadayasu and {Ebisawa}, Ken and {Eckart}, Megan and {Eckert}, Dominique and {Enoto}, Teruaki and {Eguchi}, Satoshi and {Ezoe}, Yuichiro and {Foster}, Adam and {Fujimoto}, Ryuichi and {Fujita}, Yutaka and {Fukazawa}, Yasushi and {Fukushima}, Kotaro and {Furuzawa}, Akihiro and {Gallo}, Luigi and {Garcia}, Javier A. and {Gu}, Liyi and {Guainazzi}, Matteo and {Hagino}, Kouichi and {Hamaguchi}, Kenji and {Hatsukade}, Isamu and {Hayashi}, Katsuhiro and {Hayashi}, Takayuki and {Hell}, Natalie and {Hodges-Kluck}, Edmund and {Hornschemeier}, Ann and {Ichinohe}, Yuto and {Ishida}, Manabu and {Ishikawa}, Kumi and {Ishisaki}, Yoshitaka and {Kaastra}, Jelle and {Kallman}, Timothy and {Kara}, Erin and {Katsuda}, Satoru and {Kanemaru}, Yoshiaki and {Kelley}, Richard and {Kilbourne}, Caroline and {Kitamoto}, Shunji and {Kobayashi}, Shogo and {Kohmura}, Takayoshi and {Kubota}, Aya and {Leutenegger}, Maurice and {Loewenstein}, Michael and {Maeda}, Yoshitomo and {Markevitch}, Maxim and {Matsumoto}, Hironori and {Matsushita}, Kyoko and {McCammon}, Dan and {McNamara}, Brian and {Mernier}, Francois and {Miller}, Eric D. and {Miller}, Jon M. and {Mitsuishi}, Ikuyuki and {Mizumoto}, Misaki and {Mizuno}, Tsunefumi and {Mori}, Koji and {Mukai}, Koji and {Murakami}, Hiroshi and {Mushotzky}, Richard and {Nakajima}, Hiroshi and {Nakazawa}, Kazuhiro and {Ness}, Jan-Uwe and {Nobukawa}, Kumiko and {Nobukawa}, Masayoshi and {Noda}, Hirofumi and {Odaka}, Hirokazu and {Ogawa}, Shoji and {Ogorzalek}, Anna and {Okajima}, Takashi and {Ota}, Naomi and {Paltani}, Stephane and {Petre}, Robert and {Plucinsky}, Paul and {Porter}, Frederick Scott and {Pottschmidt}, Katja and {Sato}, Kosuke and {Sato}, Toshiki and {Sawada}, Makoto and {Seta}, Hiromi and {Shidatsu}, Megumi and {Simionescu}, Aurora and {Smith}, Randall and {Suzuki}, Hiromasa and {Szymkowiak}, Andrew and {Takahashi}, Hiromitsu and {Takeo}, Mai and {Tamagawa}, Toru and {Tamura}, Keisuke and {Tanaka}, Takaaki and {Tanimoto}, Atsushi and {Tashiro}, Makoto and {Terada}, Yukikatsu and {Terashima}, Yuichi and {Tsuboi}, Yohko and {Tsujimoto}, Masahiro and {Tsunemi}, Hiroshi and {Tsuru}, Takeshi G. and {Tumer}, Aysegul and {Uchida}, Hiroyuki and {Uchida}, Nagomi and {Uchida}, Yuusuke and {Uchiyama}, Hideki and {Ueda}, Yoshihiro and {Uno}, Shinichiro and {Vink}, Jacco and {Watanabe}, Shin and {Williams}, Brian J. and {Yamada}, Satoshi and {Yamada}, Shinya and {Yamaguchi}, Hiroya and {Yamaoka}, Kazutaka and {Yamasaki}, Noriko and {Yamauchi}, Makoto and {Yamauchi}, Shigeo and {Yaqoob}, Tahir and {Yoneyama}, Tomokage and {Yoshida}, Tessei and {Yukita}, Mihoko and {Zhuravleva}, Irina and {Neilsen}, Joey and {Tomaru}, Ryota and {Mehdipour}, Missagh},
        title = "{Stratified wind from a super-Eddington X-ray binary is slower than expected}",
      journal = {arXiv e-prints},
         year = 2025,
        month = sep,
          eid = {arXiv:2509.14555},
        pages = {arXiv:2509.14555},
          doi = {10.48550/arXiv.2509.14555},
archivePrefix = {arXiv},
       eprint = {2509.14555},
 primaryClass = {astro-ph.HE},
       adsurl = {https://ui.adsabs.harvard.edu/abs/2025arXiv250914555X}
}

@ARTICLE{miller25apj,
       author = {{Miller}, Jon M. and {Mizumoto}, Misaki and {Shidatsu}, Megumi and {Ballhausen}, Ralf and {Behar}, Ehud and {D{\'\i}az Trigo}, Mar{\'\i}a and {Done}, Chris and {Dotani}, Tadayasu and {Garc{\'\i}a}, Javier A. and {Kallman}, Timothy and {Kobayashi}, Shogo B. and {Kubota}, Aya and {Smith}, Randall and {Takahashi}, Hiromitsu and {Tashiro}, Makoto and {Ueda}, Yoshihiro and {Vink}, Jacco and {Yamada}, Shinya and {Watanabe}, Shin and {Iizuka}, Ryo and {Terada}, Yukikatsu and {Baluta}, Chris and {Kanemaru}, Yoshiaki and {Ogawa}, Shoji and {Yoshida}, Tessei and {Hayashi}, Katsuhiro},
        title = "{XRISM Spectroscopy of the Stellar-mass Black Hole 4U 1630-472 in Outburst}",
      journal = {\apjl},
         year = 2025,
        month = jul,
       volume = {988},
       number = {1},
          eid = {L28},
        pages = {L28},
          doi = {10.3847/2041-8213/ade25c},
archivePrefix = {arXiv},
       eprint = {2506.07319},
 primaryClass = {astro-ph.HE},
       adsurl = {https://ui.adsabs.harvard.edu/abs/2025ApJ...988L..28M}
}

@ARTICLE{tsujimoto25pasj,
       author = {{Tsujimoto}, Masahiro and {Enoto}, Teruaki and {D{\'\i}az Trigo}, Mar{\'\i}a and {Hell}, Natalie and {Chakraborty}, Priyanka and {Leutenegger}, Maurice A. and {Loewenstein}, Michael and {Pradhan}, Pragati and {Shidatsu}, Megumi and {Takahashi}, Hiromitsu and {Yaqoob}, Tahir},
        title = "{Outflowing photoionized plasma in Circinus X-1 using the high-resolution X-ray spectrometer Resolve onboard XRISM and the radiative transfer code cloudy}",
      journal = {\pasj},
         year = 2025,
        month = apr,
          doi = {10.1093/pasj/psaf022},
archivePrefix = {arXiv},
       eprint = {2503.08254},
 primaryClass = {astro-ph.HE},
       adsurl = {https://ui.adsabs.harvard.edu/abs/2025PASJ..tmp...30T}
}

@ARTICLE{Kaastra2002,
       author = {{Kaastra}, J.~S. and {Steenbrugge}, K.~C. and {Raassen}, A.~J.~J. and {van der Meer}, R.~L.~J. and {Brinkman}, A.~C. and {Liedahl}, D.~A. and {Behar}, E. and {de Rosa}, A.},
        title = "{X-ray spectroscopy of NGC 5548}",
      journal = {\aap},
         year = 2002,
        month = may,
       volume = {386},
        pages = {427-445},
          doi = {10.1051/0004-6361:20020235},
archivePrefix = {arXiv},
       eprint = {astro-ph/0202481},
 primaryClass = {astro-ph},
       adsurl = {https://ui.adsabs.harvard.edu/abs/2002A&A...386..427K}
}

@ARTICLE{wachter05apj,
       author = {{Wachter}, Stefanie and {Wellhouse}, Joseph W. and {Patel}, Sandeep K. and {Smale}, Alan P. and {Alves}, Joao F. and {Bouchet}, Patrice},
        title = "{Chandra HRC Localization of the Low-Mass X-Ray Binaries X1624-490 and X1702-429: The Infrared Counterparts}",
      journal = {\apj},
         year = 2005,
        month = mar,
       volume = {621},
       number = {1},
        pages = {393-397},
          doi = {10.1086/427407},
archivePrefix = {arXiv},
       eprint = {astro-ph/0411360},
 primaryClass = {astro-ph},
       adsurl = {https://ui.adsabs.harvard.edu/abs/2005ApJ...621..393W}
}

@ARTICLE{Tashiro25,
       author = {{Tashiro}, Makoto and {Kelley}, Richard and {Watanabe}, Shin and {Maejima}, Hironori and {Reichenthal}, Lillian and {Toda}, Kenichi and {Hartz}, Leslie and {Santovincenzo}, Andrea and {Matsushita}, Kyoko and {Yamaguchi}, Hiroya and {Petre}, Robert and {Williams}, Brian and {Guainazzi}, Matteo and {Costantini}, Elisa and {Takei}, Yoh and {Ishisaki}, Yoshitaka and {Fujimoto}, Ryuichi and {Henegar-Leon}, Joy and {Sneiderman}, Gary and {Tomida}, Hiroshi and {Mori}, Koji and {Nakajima}, Hiroshi and {Terada}, Yukikatsu and {Holland}, Matthew and {Loewenstein}, Michael and {Miller}, Eric and {Sawada}, Makoto and {Kallman}, Timothy and {Kaastra}, Jelle and {Done}, Chris and {Enoto}, Teruaki and {Bamba}, Aya and {Corrales}, Lia and {Ueda}, Yoshihiro and {Kara}, Erin and {Zhuravleva}, Irina and {Fujita}, Yutaka and {Arai}, Yoshitaka and {Audard}, Marc and {Awaki}, Hisamitsu and {Ballhausen}, Ralf and {Baluta}, Chris and {Bando}, Nobutaka and {Behar}, Ehud and {Bialas}, Thomas and {Boissay-Malaquin}, Rozenn and {Brenneman}, Laura and {Brown}, Gregory V. and {Chiao}, Meng and {Cumbee}, Renata and {de Vries}, Cor and {den Herder}, Jan-Willem and {D{\'\i}az Trigo}, Mar{\'\i}a and {DiPirro}, Michael and {Dotani}, Tadayasu and {Carrero}, Jacobo Ebrero and {Ebisawa}, Ken and {Eckart}, Megan and {Eckert}, Dominique and {Eguchi}, Satoshi and {Ezoe}, Yuichiro and {Ferrigno}, Carlo and {Foster}, Adam and {Fukazawa}, Yasushi and {Fukushima}, Kotaro and {Furuzawa}, Akihiro and {Gallo}, Luigi and {Garcia Martinez}, Javier and {Gorter}, Nathalie and {Grim}, Martin and {Gu}, Liyi and {Hagino}, Kouichi and {Hamaguchi}, Kenji and {Hatsukade}, Isamu and {Hayashi}, Katsuhiro and {Hayashi}, Takayuki and {Hell}, Natalie and {Hodges-Kluck}, Edmund and {Horiuchi}, Takafumi and {Hornschemeier}, Ann and {Hoshino}, Akio and {Ichinohe}, Yuto and {Ikuta}, Chisato and {Iizuka}, Ryo and {Ishi}, Daiki and {Ishida}, Manabu and {Ishihama}, Naoki and {Ishikawa}, Kumi and {Ishimura}, Kosei and {Jaffe}, Tess and {Katsuda}, Satoru and {Kanemaru}, Yoshiaki and {Kenyon}, Steven and {Kilbourne}, Caroline and {Kimball}, Mark and {Kitamoto}, Shunji and {Kobayashi}, Shogo and {Kohmura}, Takayoshi and {Kubota}, Aya and {Leutenegger}, Maurice and {Maeda}, Yoshitomo and {Markevitch}, Maxim and {Matsumoto}, Hironori and {Matsuzaki}, Keiichi and {McCammon}, Dan and {McLaughlin}, Brian and {McNamara}, Brian and {Mernier}, Francois and {Miko}, Joseph and {Miller}, Jon and {Minesugi}, Kenji and {Mitani}, Shinji and {Mitsuishi}, Ikuyuki and {Mizumoto}, Misaki and {Mizuno}, Tsunefumi and {Mukai}, Koji and {Murakami}, Hiroshi and {Mushotzky}, Richard and {Nakazawa}, Kazuhiro and {Natsukari}, Chikara and {Ness}, Jan-Uwe and {Nigo}, Kenichiro and {Nishiyama}, Mari and {Nobukawa}, Kumiko and {Nobukawa}, Masayoshi and {Noda}, Hirofumi and {Odaka}, Hirokazu and {Ogawa}, Mina and {Ogawa}, Shoji and {Ogorzalek}, Anna and {Okajima}, Takashi and {Okamoto}, Atsushi and {Ota}, Naomi and {Ozaki}, Masanobu and {Paltani}, Stephane and {Plucinsky}, Paul and {Porter}, F. Scott and {Pottschmidt}, Katja and {Quero}, Jose Antonio and {Sasaki}, Takahiro and {Sato}, Kosuke and {Sato}, Rie and {Sato}, Toshiki and {Sato}, Yoichi and {Seta}, Hiromi and {Shida}, Maki and {Shidatsu}, Megumi and {Shigeto}, Shuhei and {Shipman}, Russel and {Shinozaki}, Keisuke and {Shirron}, Peter and {Simionescu}, Aurora and {Smith}, Randall and {Soong}, Yang and {Suzuki}, Hiromasa and {Szymkowiak}, Andrew and {Takahashi}, Hiromitsu and {Takeo}, Mai and {Tamagawa}, Toru and {Tamura}, Keisuke and {Tanaka}, Takaaki and {Tanimoto}, Atsushi and {Terashima}, Yuichi and {Tsuboi}, Yohko and {Tsujimoto}, Masahiro and {Tsunemi}, Hiroshi and {Tsuru}, Takeshi and {Uchida}, Hiroyuki and {Uchida}, Nagomi and {Uchida}, Yuusuke and {Uchiyama}, Hideki and {Uno}, Shinichiro and {Vink}, Jacco and {Witthoeft}, Michael and {Wolfs}, Rob and {Yamada}, Satoshi and {Yamada}, Shinya and {Yamaoka}, Kazutaka and {Yamasaki}, Noriko and {Yamauchi}, Makoto and {Yamauchi}, Shigeo and {Yanagase}, Keiichi and {Yaqoob}, Tahir and {Yasuda}, Susumu and {Yoneyama}, Tomokage and {Yoshida}, Tessei and {Yukita}, Miohoko},
        title = "{X-Ray Imaging and Spectroscopy Mission}",
      journal = {\pasj},
         year = 2025,
        month = apr,
          doi = {10.1093/pasj/psaf023},
       adsurl = {https://ui.adsabs.harvard.edu/abs/2025PASJ..tmp...28T}
}

@ARTICLE{Mehdipour16,
       author = {{Mehdipour}, M. and {Kaastra}, J.~S. and {Kallman}, T.},
        title = "{Systematic comparison of photoionised plasma codes with application to spectroscopic studies of AGN in X-rays}",
      journal = {\aap},
         year = 2016,
        month = dec,
       volume = {596},
          eid = {A65},
        pages = {A65},
          doi = {10.1051/0004-6361/201628721},
archivePrefix = {arXiv},
       eprint = {1610.03080},
 primaryClass = {astro-ph.HE},
       adsurl = {https://ui.adsabs.harvard.edu/abs/2016A&A...596A..65M}
}

@ARTICLE{mroz2019ApJ,
       author = {{Mr{\'o}z}, Przemek and {Udalski}, Andrzej and {Skowron}, Dorota M. and {Skowron}, Jan and {Soszy{\'n}ski}, Igor and {Pietrukowicz}, Pawe{\l} and {Szyma{\'n}ski}, Micha{\l} K. and {Poleski}, Rados{\l}aw and {Koz{\l}owski}, Szymon and {Ulaczyk}, Krzysztof},
        title = "{Rotation Curve of the Milky Way from Classical Cepheids}",
      journal = {\apjl},
         year = 2019,
        month = jan,
       volume = {870},
       number = {1},
          eid = {L10},
        pages = {L10},
          doi = {10.3847/2041-8213/aaf73f},
archivePrefix = {arXiv},
       eprint = {1810.02131},
 primaryClass = {astro-ph.GA},
       adsurl = {https://ui.adsabs.harvard.edu/abs/2019ApJ...870L..10M}
}

@ARTICLE{kilbourne2018,
       author = {{Kilbourne}, Caroline A. and {Sawada}, Makoto and {Tsujimoto}, Masahiro and {Angellini}, Lorella and {Boyce}, Kevin R. and {Eckart}, Megan E. and {Fujimoto}, Ryuichi and {Ishisaki}, Yoshitaka and {Kelley}, Richard L. and {Koyama}, Shu and {Leutenegger}, Maurice A. and {Loewenstein}, Michael and {McCammon}, Dan and {Mitsuda}, Kazuhisa and {Nakashima}, Shinya and {Porter}, Frederick S. and {Seta}, Hiromi and {Takei}, Yoh and {Tashiro}, Makoto S. and {Terada}, Yukikatsu and {Yamada}, Shinya and {Yamasaki}, Noriko Y.},
        title = "{In-flight calibration of Hitomi Soft X-ray Spectrometer. (1) Background}",
      journal = {\pasj},
         year = 2018,
        month = mar,
       volume = {70},
       number = {2},
          eid = {18},
        pages = {18},
          doi = {10.1093/pasj/psx139},
       adsurl = {https://ui.adsabs.harvard.edu/abs/2018PASJ...70...18K}
}

@ARTICLE{mdarias25,
       author = {{Mu{\~n}oz-Darias}, Teo and {D{\'\i}az Trigo}, Mar{\'\i}a and {Done}, Chris and {Ponti}, Gabriele and {Tomaru}, Ryota},
        title = "{Accretion disc winds in X-ray binaries}",
      journal = {arXiv e-prints},
         year = 2026,
        month = jan,
          eid = {arXiv:2601.05319},
        pages = {arXiv:2601.05319},
archivePrefix = {arXiv},
       eprint = {2601.05319},
 primaryClass = {astro-ph.HE},
       adsurl = {https://ui.adsabs.harvard.edu/abs/2026arXiv260105319M}
}

@ARTICLE{porter25,
       author = {{Porter}, Frederick Scott and {Kilbourne}, Caroline A. and {Chiao}, Meng P. and {Cumbee}, Renata S. and {Eckart}, Megan E. and {Fujimoto}, Ryuichi and {Ishisaki}, Yoshitaka and {Kanemaru}, Yoshiaki and {Kelley}, Richard L. and {Leutenegger}, Maurice Andrew and {Maeda}, Yoshitomo and {Mizumoto}, Misaki and {Sato}, Kosuke and {Sawada}, Makoto and {Sneiderman}, Gary A. and {Takei}, Yoh and {Tsujimoto}, Masahiro and {Uchida}, Yuusuke and {Watanabe}, Tomomi and {Yamada}, Shin’ya},
    doi = {https://doi.org/10.1117/1.JATIS.11.4.042016},     
    title = "{In-flight performance of the XRISM/Resolve detector system}",
      journal = {Journal of Astronomical Telescopes, Instruments, and Systems},
    volume = {11},
       number = {4},
eid = {042016},
        pages = {042016},
    year= 2025,

}

@ARTICLE{tomaru20mnras,
       author = {{Tomaru}, Ryota and {Done}, Chris and {Ohsuga}, Ken and {Odaka}, Hirokazu and {Takahashi}, Tadayuki},
        title = "{The thermal-radiative wind in the neutron star low-mass X-ray binary GX 13 + 1}",
      journal = {\mnras},
         year = 2020,
        month = oct,
       volume = {497},
       number = {4},
        pages = {4970-4980},
          doi = {10.1093/mnras/staa2254},
archivePrefix = {arXiv},
       eprint = {2007.14607},
 primaryClass = {astro-ph.HE},
       adsurl = {https://ui.adsabs.harvard.edu/abs/2020MNRAS.497.4970T}
}

@ARTICLE{trueba20apj,
       author = {{Trueba}, Nicolas and {Miller}, J.~M. and {Fabian}, A.~C. and {Kaastra}, J. and {Kallman}, T. and {Lohfink}, A. and {Proga}, D. and {Raymond}, J. and {Reynolds}, C. and {Reynolds}, M. and {Zoghbi}, A.},
        title = "{A Redshifted Inner Disk Atmosphere and Transient Absorbers in the Ultracompact Neutron Star X-Ray Binary 4U 1916-053}",
      journal = {\apjl},
         year = 2020,
        month = aug,
       volume = {899},
       number = {1},
          eid = {L16},
        pages = {L16},
          doi = {10.3847/2041-8213/aba9de},
archivePrefix = {arXiv},
       eprint = {2008.01083},
 primaryClass = {astro-ph.HE},
       adsurl = {https://ui.adsabs.harvard.edu/abs/2020ApJ...899L..16T}
}

\end{document}